\documentclass[twocolumn,showpacs,amsmath,amssymb,floatfix]{revtex4}

\usepackage{graphicx}

\newcommand{\ii}{\mathrm{i}}
\newcommand{\hc}{\mathrm{h.c.}}

\newcommand{\vi}{{\pmb i}}

\newcommand{\si}{\hat{s}_{{\pmb i}}}
\newcommand{\sid}{\hat{s}^{\dagger}_{{\pmb i}}}
\newcommand{\tialpha}{\hat{t}_{{\pmb i}\alpha}}
\newcommand{\tialphad}{\hat{t}^{\dagger}_{{\pmb i}\alpha}}

\newcommand{\tibetad}{\hat{t}^{\dagger}_{{\pmb i}\beta}}
\newcommand{\tigamma}{\hat{t}_{{\pmb i}\gamma}}

\newcommand{\sj}{\hat{s}_{{\pmb j}}}
\newcommand{\sjd}{\hat{s}^{\dagger}_{{\pmb j}}}
\newcommand{\tjalpha}{\hat{t}_{{\pmb j}\alpha}}
\newcommand{\tjalphad}{\hat{t}^{\dagger}_{{\pmb j}\alpha}}
\newcommand{\tjbeta}{\hat{t}_{{\pmb j}\beta}}
\newcommand{\tjbetad}{\hat{t}^{\dagger}_{{\pmb j}\beta}}
\newcommand{\tjgamma}{\hat{t}_{{\pmb j}\gamma}}
\newcommand{\tjgammad}{\hat{t}^{\dagger}_{{\pmb j}\gamma}}

\begin{document}
\title{Single hole dynamics in the Kondo Necklace and Bilayer Heisenberg  models
on a square lattice.}
\author{C. Br\"unger and F.F. Assaad}

\affiliation{ Institut f\"ur Theoretische Physik und Astrophysik,
Universit\"at W\"urzburg, Am Hubland, D-97074 W\"urzburg, Germany }

\begin{abstract}
We study single hole dynamics in the bilayer Heisenberg and Kondo Necklace models.
Those models exhibit a magnetic order-disorder quantum phase transition as a function of 
the interlayer  coupling $J_{\bot}$. At strong coupling in the disordered phase, 
both models have a single-hole dispersion relation with band maximum at ${\pmb p} = (\pi,\pi)$ 
and an effective  mass at this ${\pmb p}-$point which scales as the hopping matrix 
element $t$.  In the  Kondo Necklace model, we show that the effective mass at 
${\pmb p} = (\pi,\pi)$  
remains finite for all considered values of $J_{\bot}$ such that the strong coupling 
features  of the dispersion relation are apparent  down to weak coupling.  In contrast, in 
the bilayer Heisenberg model, the effective mass diverges at a finite  value of $J_{\bot}$. 
This divergence of the effective mass is unrelated to the magnetic quantum phase transition and 
at weak coupling the dispersion  relation maps onto that of a single hole doped in a planar 
antiferromagnet with band maximum at ${\pmb p} = (\pi/2,\pi/2)$. 
We equally study the behavior of the quasiparticle residue in the vicinity of the 
magnetic quantum phase transition both for a mobile and static hole.  In contrast to 
analytical approaches, our  numerical results do not unambiguously support the 
fact that the  quasiparticle residue of the static hole vanishes in the vicinity of 
the critical point.   The above  results are obtained  with a generalized version of 
the loop algorithm to include single hole dynamics on lattice sizes up to $20 \times 20$. 
\end{abstract}


\maketitle

\section{Introduction}

The modeling of heavy fermion systems is based on an array of localized spin degrees 
of freedom coupled antiferromagnetically to conduction electrons.  Those models 
show competing interactions which lead to magnetic quantum phase transitions as  
a function of the  antiferromagnetic exchange interaction $J$. 
Kondo screening of the localized spins, dominant at large $J$,  favors a 
paramagnetic heavy fermion ground state, where the localized spins participate in 
the Luttinger volume. In contrast, the RKKY interaction  favors  magnetic ordering and 
is dominant at small  values of $J$.  There has recently been renewed  
interest concerning the understanding this quantum phase transition. In particular,  recent  
Hall experiments \cite{Paschen04} suggest the interpretation that in the vicinity of the 
quantum phase transition
the localized spins drop out of the Luttinger volume.  Starting from the 
paramagnetic phase, this  transition from a large to small 
Fermi surface should coincide with a effective mass divergence of the heavy fermion band.

Motivated by the above, we consider here a very simplified situation  namely that of 
a doped hole in the Kondo insulating state as realized  by the Kondo necklace 
and related models. Although this is not of direct relevance for the study of the Fermi surface, 
it does allow us to investigate  the form of the quasiparticle  dispersion relation from 
strong to weak coupling for a variety of models.  Our aim here is two fold. On one hand 
we address the question of the divergence of the effective mass as a function 
of coupling  for different models, and on the other hand the fate of the quasiparticle 
residue in the vicinity of the quantum phase transition.

The KLM emerges from the periodic Anderson model (PAM), where we have localized orbitals 
(LO) with on-site Hubbard interaction $U_{f}$ and extended orbitals (EO), which form a 
conduction band with dispersion 
$\varepsilon({\pmb p})=-2t\left(\cos p_{x} + \cos p_{y}\right)$. 
The overlap between the LOs and the EOs within each unit cell is described by the 
hybridization matrix element $V$. For large $U_{f}$ charge fluctuations  on the localized 
orbitals becomes negligible and the PAM maps   via the Schrieffer-Wolff transformation onto 
the KLM \cite{Schrieffer66,Tsunetsugu97}:
\begin{eqnarray}
\label{KLM}
\hat{H}_{\text{KLM}}=\sum_{{\pmb p},\sigma}\varepsilon ({\pmb p})
\hat{c}^{\dagger}_{{\pmb p}\sigma}\hat{c}_{{\pmb p}\sigma}
+\mathcal{J}\sum_{i}\hat{{\pmb S}}^{c}_{{\pmb i}}\hat{{\pmb S}}^{f}_{{\pmb i}}\,\, .
\end{eqnarray}
Here $\hat{{\pmb S}}^{c}_{{\pmb i}}$ and $\hat{{\pmb S}}^{f}_{{\pmb i}}$ are spin $1/2$ 
operators for the extended orbitals and the localized orbitals respectively. 
In the first term, 
which represents the hopping processes, the fermionic operators 
$\hat{c}^{\dagger}_{{\pmb p}\sigma}$ ($\hat{c}_{{\pmb p}\sigma}$ ) create (annihilate) 
electrons in the conduction band with wave vector ${\pmb p}$ and $z$-component of  spin 
$\sigma$. At half-filling -- one conduction electron per localized spin --
the two-dimensional KLM is an insulator
and  shows a magnetic order-disorder quantum 
phase transition at a critical value of $\mathcal{J}_{c}/t=1.45\pm 0.05$ \cite{Capponi00}. 

\begin{figure}%[h]
\centering
\includegraphics[height=3cm]{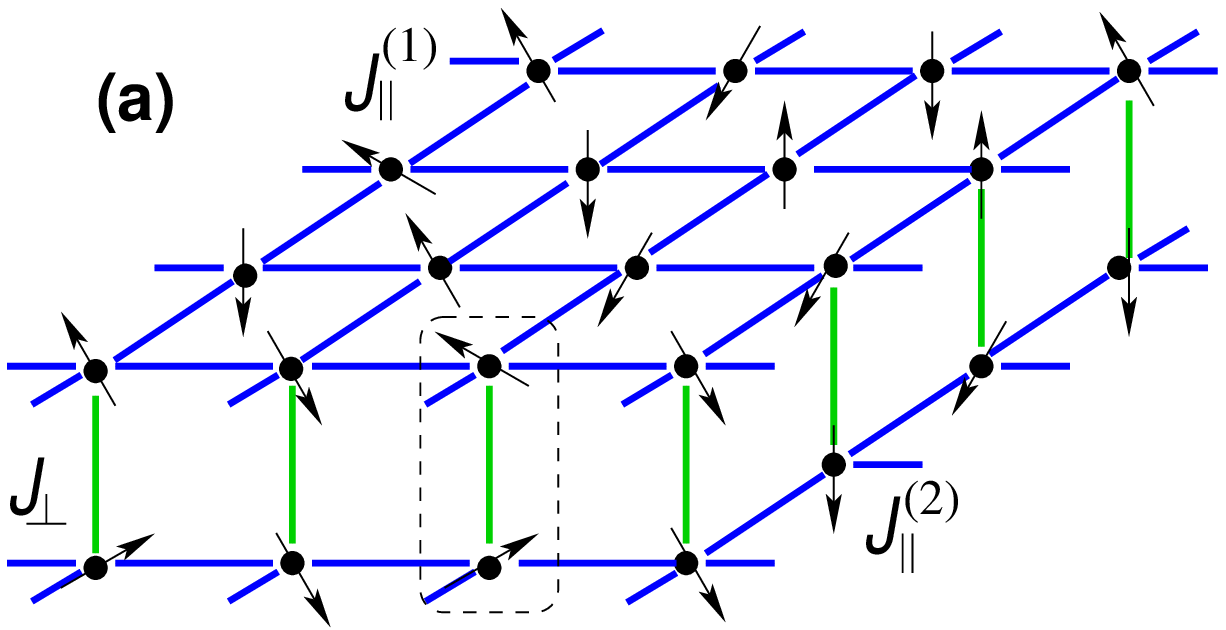}
\includegraphics[height=3cm]{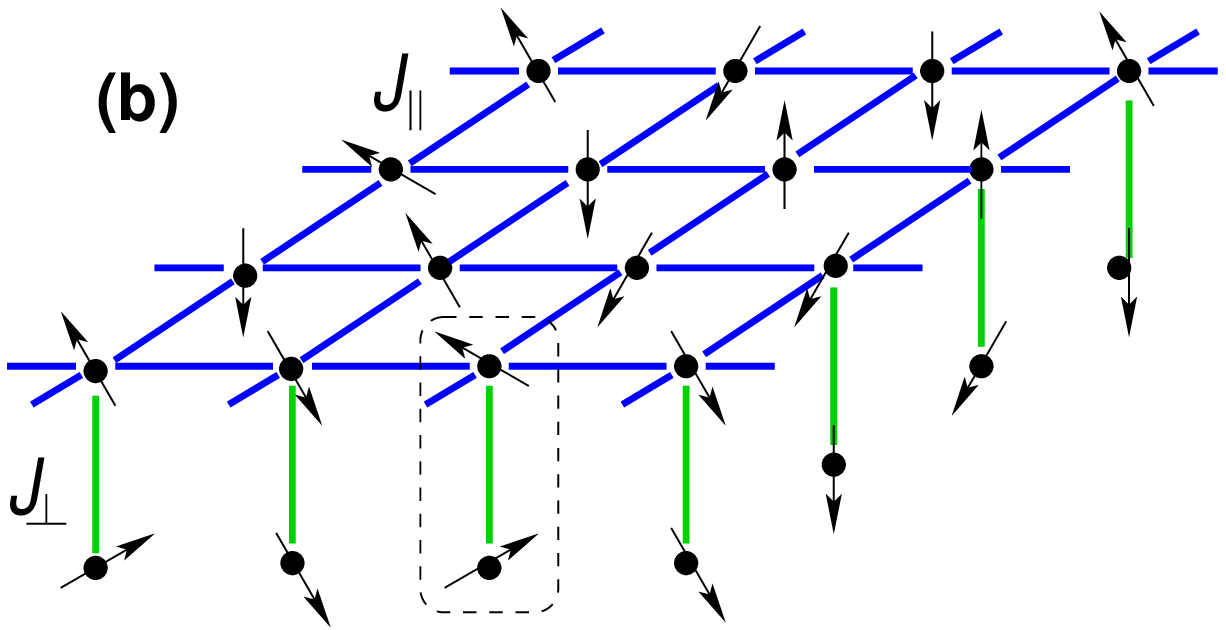}
\caption{\label{lattice}{\small (a) Isotropic bilayer Heisenberg model. 
(b) Kondo Necklace model, that is related to the $U$KLM. In both cases the system dimerizes 
for large $J_{\bot}$, so that the AF ordering breaks down.}}
\end{figure}

By taking into account an additional  Coulomb repulsion $U$ between electrons within the 
conduction band, one obtains a modification of the KLM, the $U$KLM:

\begin{eqnarray}
\hat{H}_{U\text{KLM}}
&=&\sum_{{\pmb p},\sigma}\varepsilon ({\pmb p})
   \hat{c}^{\dagger}_{{\pmb p}\sigma}\hat{c}_{{\pmb p}\sigma}
   +\mathcal{J}\sum_{i}\hat{{\pmb S}}^{c}_{{\pmb i}}\hat{{\pmb S}}^{f}_{{\pmb i}}
   \nonumber\\
& &+U\sum_{{\pmb i}}\left(\hat{n}_{{\pmb i}\uparrow}-\tfrac{1}{2}\right)
   \left(\hat{n}_{{\pmb i}\downarrow}-\tfrac{1}{2}\right)\,\, .
\end{eqnarray}
Here, 
$\hat{n}_{{\pmb i}\sigma}=\hat{c}^{\dagger}_{{\pmb i}\sigma}\hat{c}_{{\pmb i}\sigma}$ is the 
density operator for electrons with spin $\sigma$ in the conduction band. 
The additional Coulomb repulsion 
displaces the quantum critical point towards smaller value of $\mathcal{J}_{c}/t$. 
However the  physics, in particular  the single hole dynamics,  remains unchange  
\cite{Feldbacher02}.  This allows us to take the limit $U/t \rightarrow \infty $ to 
map the UKLM onto a Kondo necklace model (KNM) which we write as:
\begin{eqnarray}
\label{BHM_model}
\hat{H}
& = & J_{\bot}\sum_{{\pmb i}}\hat{{\pmb S}}^{(1)}_{{\pmb i}}\hat{{\pmb S}}^{(2)}_{{\pmb i}} +
      \sum_{\langle{\pmb i}{\pmb j}\rangle}\sum_{m}J^{(m)}_{\|}
      \hat{{\pmb S}}^{(m)}_{{\pmb i}}\hat{{\pmb S}}^{(m)}_{{\pmb j}}.
      \label{starth}
\end{eqnarray}
Here $\hat{{\pmb S}}^{(m)}_{{\pmb i}}$ is a spin $1/2$ operator, 
which acts on a spin degree of freedom at site ${\pmb i}$.
$J^{(m)}_{\|}$ stands for the intralayer exchange and the upper index $m=1,2$ labels the 
two different layers.
The interlayer exchange, formerly the AF coupling $\mathcal{J}$ between LOs and EOs, 
is now characterized by $J_{\bot}$. 
Clearly, since we have motivated the  KNM from a strong coupling limit of the UKLM, we 
have to set: 
\begin{equation}
J^{(1)}_{\|}\equiv J_{\|} \quad J^{(2)}_{\|}=0 
\quad\text{for the KNM.}
\end{equation}

The above models all have in common that the only interaction between the localized spins 
stems from the RKKY interaction.   This in turn  leads to the fact that at $\mathcal{J} =0$ for 
the KLM and UKLM or $J_{\bot} =0$ for the KNM  the ground state is  macroscopically degenerate. 
To lift the {\it pathology} we finally consider a Bilayer Heisenberg Model (BHM)  in which 
an independent exchange  between the localized spins is  explicitly included   in the
 Hamiltonian.   Hence we will equally consider an  Isotropic BHM  which takes the form  of Eq. 
(\ref{BHM_model})  with:
\begin{equation}
J^{(1)}_{\|}=J^{(2)}_{\|}\equiv J_{\|}
\quad\text{for the  isotropic BHM.}
\end{equation}
Both the KNM and BHM systems are sketched in FIG. \ref{lattice}.

The main results and organization of the paper are the following. 
In section \ref{methods} we give a short overview of the quantum Monte Carlo (QMC) 
method.  We use a generalization of the loop algorithm  which allows  for the 
calculation of the imaginary time Green's function of the doped hole \cite{Brunner00}.  Dynamical 
information is obtained with a  stochastic  Maximum Entropy method \cite{Beach04,Sandvik98}. 
In the first part of section \ref{dynamics} we present our results for 
the spin dynamics. This includes the determination of the quantum critical point for 
the isotropic BHM as well as the Kondo Necklace model (KNM) by QMC methods.  
In the second part of that section we analyze the single 
particle spectral function. It turns out, that there are significant differences 
between the models. We can identify two classes of models: In the isotropic BHM the 
dispersion is continously deformed with decreasing interplanar coupling 
$J_{\bot}/J_{\|}$ resulting in 
a displacement of the maximum from 
${\pmb p}=(\pi,\pi)$ to ${\pmb p}=(\tfrac{\pi}{2},\tfrac{\pi}{2})$.  In other words, 
the effective mass -- as defined by the inverse curvature of the quasiparticle 
dispersion relation --  at ${\pmb p}=(\pi,\pi)$ diverges  at a finite  value of the interplanar
coupling.  This divergence of the effective  mass is not related  to the magnetic order-disorder 
transition.  
In contrast, in the KLM related models, UKLM and KNM,  the maximum of the quasiparticle 
dispersion relation is pinned at ${\pmb p}=(\pi,\pi)$  irrespective of the 
value of the interplanar coupling. In those models the effective mass  at ${\pmb p}=(\pi,\pi)$ 
grows as a function of decreasing interplanar coupling, but remains finite. 

In section \ref{qpr} we turn to the analysis of the quasi particle residue 
(QPR) across the quantum phase transition. 
To gain intuition, 
we first carry out an approximate calculation in the lines of Ref. \cite{Sushkov00}. 
The physics of the spin 
system may be solved in the framework of a bond  mean-field calculation. Here, the 
disordered phase is described in terms of a condensate of singlets between the planes 
and gaped spin 1 excitations (magnons). At the critical point the magnons condense at 
the AF wave vector thus  generating the static antiferromagnetic order. Within this 
framework one can compute the coupling of the mobile hole with the magnetic 
fluctuations and study the hole dynamics within a self-consistent Born approximation. 
The result of the calculation shows that the quasiparticle weight at wave vectors on 
the magnetic Brillouin zone [$\epsilon({\pmb p}) =  \epsilon({\pmb p} + {\pmb Q}) $ with 
$\pmb{Q} = (\pi,\pi)$ ] vanish as the square root of the spin gap. In contrast 
the QMC determination of the quasiparticle residue on lattices up to $20 \times 20$ 
for static and dynamical holes does not unambiguously  support this point of view. 

\section{Numerical Methods}
\label{methods}
We  use the world line   QMC method with loop updates \cite{Evertz97} to investigate the 
physics of the BHM and KNM.  To investigate the spin dynamics we compute both the 
spin stiffness as well as the dynamical spin structure factor. Our analysis 
of the single hole dynamics is based on the calculation of the imaginary time Green's function. 
Analytical continuation with the use of the stochastic Maxent Method provides the 
spectral function and the quasiparticle residue is extracted from the asymptotic 
behavior of the imaginary time Green's function.   Below, we discuss in more details the 
calculation of each observables.

\subsubsection*{Spin Stiffness}
To probe for long-ranged magnetic order 
we introduce a continuous twist in spin space which, when cumulated along the 
length $L$ along  (e.g.) the $x$-axis, amounts to a twist of angle $\phi$ around a certain 
spin axis ${\pmb e}$. This means thus the boundary conditions read: 
$\hat{{\pmb S}}_{{\pmb i}+L{\pmb e}_{x}}=R\left[{\pmb e},\phi\right]\hat{{\pmb S}}_{{\pmb i}}$, 
where $R\left[{\pmb e},\phi\right]$ is a matrix describing an $SO(3)$ rotation 
around the axis ${\pmb e}$ 
by the angle $\phi$. The spin stiffness is  then defined as
\begin{eqnarray}
\rho_{s} = \left. -\frac{1}{L^{d-2}}\frac{1}{\beta}\frac{\partial^{2}}{\partial
\phi^{2}}\ln Z(\phi)  \right|_{\phi=0}
\end{eqnarray}
with $\beta$ as inverse temperature, $L$ as the linear size of the system, $d$ 
the dimensionality and $ Z(\phi) $ the twist dependent  partition function.  
In the presence of long-range order $\rho_{s}$  takes a finite value and in a
disordered phase it vanishes. 

Within the world-line algorithm, the spin stiffness is related to the winding number 
$\mathcal{W}_{x}$ of the world 
line configurations along the axis of cumulatively twisted spins (e.g. $x$-axis).
In particular, in the limit $\Delta\tau\to 0$ it takes the simple form
\begin{eqnarray}
\rho_{s}=\frac{1}{L^{d}}\frac{1}{\beta}\mathcal{W}^{2}_{x}.
\end{eqnarray}
\subsubsection*{Spin Correlations}
Within the QMC it is easy to obtain the spin correlations 
$\langle S^{z}_{{\pmb i}}(\tau)S^{z}_{{\pmb j}}(0)\rangle$ in real space and imaginary time $\tau$, 
where the imaginary time evolution of the spin operator reads  
$S^{z}_{{\pmb q}}(\tau)=e^{\tau\hat{H}}S^{z}_{{\pmb q}}e^{-\tau\hat{H}}$.
Its representation in momentum space is related to the dynamical spin 
susceptibility $S({\pmb q},\omega)$ via:
\begin{eqnarray}
\langle S^{z}_{{\pmb q}}(\tau)S^{z}_{-{\pmb q}}(0)\rangle = \frac{1}{\pi}
\int d\omega e^{-\tau\omega}S({\pmb q},\omega).
\end{eqnarray}
By using the Stochastic Maximum Entropy (ME) method \cite{Beach04} 
we can  extract the dynamical spin susceptibility. 
For large $\tau$ the spin correlation function is dominated by the lowest 
excitation:
\begin{eqnarray}
\underset{\tau\rightarrow\infty}{\lim} 
\langle S^{z}_{{\pmb q}}(\tau)S^{z}_{-{\pmb q}}(0)\rangle
\propto e^{-\Omega({\pmb q})\tau}
\end{eqnarray}
where $\Omega({\pmb q})$ stands for momentum dependent gap to the first 
spin excitation. 
Thus, we obtain the gap energy $\Delta$ from the asymptotic behavior of the spin 
correlations: $\Delta \equiv \min \left[ \Omega({\pmb q}) \right]$.
\subsubsection*{ The Green's Function}
To incorporate the dynamics of a single hole into the KNM and BHM,
we consider the $tJ$-model
\begin{eqnarray}
\hat{H}_{tJ}&=&
\mathcal{P}_{S}\Big{[}- \sum_{\langle {\pmb i}{\pmb j} \rangle,\sigma} t_{\pmb{i}\pmb{j}} 
       (\hat{c}^{\dagger}_{{\pmb i}\sigma}\hat{c}_{{\pmb j}\sigma}
+\hat{c}^{\dagger}_{{\pmb j}\sigma}\hat{c}_{{\pmb i}\sigma})
\nonumber\\
&&+\displaystyle\sum_{\langle {\pmb i}{\pmb j}  \rangle} 
        J_{\pmb{i}\pmb{j}}\big{\{}\hat{{\pmb S}}_{{\pmb i}}\hat{{\pmb S}}_{{\pmb j}}
-\tfrac{1}{4}\hat{n}_{{\pmb i}}\hat{n}_{{\pmb j}}\big{\}}\Big{]}\mathcal{P}_{S}
\label{tj}
\end{eqnarray}
which describes the more general case of arbitrary filling.  Here, $ {\pmb i} $ and ${\pmb j}$
 denote  lattice sites of the bilayer  BHM, $t_{{\pmb i}{\pmb j}}$ the hopping amplitude, 
$J_{{\pmb i}{\pmb j}} $ the exchange, 
$\hat{n}_{{\pmb j}}=\hat{c}^{\dagger}_{{\pmb i}\sigma}\hat{c}_{{\pmb i}\sigma}$, 
and the sums run over nearest inter- and intraplane 
neighbors.
Finally  $\mathcal{P}_{S}$ is a projection operator onto the subspace 
$S$ with no double occupation. 
We apply a mapping, introduced
by Angelucci \cite{Angelucci95}, which separates the spin degree of freedom and 
the occupation number. 
\begin{eqnarray}
\begin{array}{lcrclcl}
|\uparrow\rangle & \longrightarrow & |1,\Uparrow\rangle &\hphantom{xxx}&
\hat{c}_{{\pmb i}\uparrow} & \longrightarrow &
\hat{\sigma}^{z,+}_{{\pmb i}}\hat{f}^{\dagger}_{{\pmb i}}-\hat{\sigma}^{z,-}_{{\pmb i}}\hat{f}_{{\pmb i}}\vphantom{\frac{1}{1}}\\
|\downarrow\rangle & \longrightarrow & |1,\Downarrow\rangle &\hphantom{xxx}&
\hat{c}^{\dagger}_{{\pmb i}\uparrow} & \longrightarrow &
\hat{\sigma}^{z,+}_{{\pmb i}}\hat{f}_{{\pmb i}}-\hat{\sigma}^{z,-}_{{\pmb i}}\hat{f}^{\dagger}_{{\pmb i}}\vphantom{\frac{1}{1}}\\
|0\rangle & \longrightarrow & |0,\Uparrow\rangle &\hphantom{xxx}&
\hat{c}_{{\pmb i}\downarrow} & \longrightarrow & (\hat{f}_{{\pmb i}}+\hat{f}^{\dagger}_{{\pmb i}})\hat{\sigma}^{+}_{{\pmb i}}
\vphantom{\frac{1}{1}}\\
|\uparrow\downarrow\rangle & \longrightarrow & |0,\Downarrow\rangle &\hphantom{xxx}&
\hat{c}^{\dagger}_{{\pmb i}\downarrow} & \longrightarrow & \hat{\sigma}^{-}_{{\pmb i}}(\hat{f}^{\dagger}_{{\pmb i}}+\hat{f}_{{\pmb i}})
\vphantom{\frac{1}{1}}
\end{array}
\end{eqnarray}
$\hat{f}^{\dagger}_{{\pmb i}}$ and $\hat{f}_{{\pmb i}}$ are spinless fermion operators which 
act on the charge degree of freedom and create (annihilate) 
a hole at site $i$: $\hat{f}^{\dagger}_{{\pmb i}}|1,\sigma\rangle  =  |0,\sigma\rangle$,  
$\hat{\sigma}^{\pm}_{{\pmb i}}$ are ladder operators for the spin degree of freedom and 
$\hat{\sigma}^{z,\pm}_{{\pmb i}}=\frac{1}{2}(1\pm\hat{\sigma}^{z}_{{\pmb i}})$ are projector 
operators acting on the spin degree of freedom. 
Within this base the Hamilton of the $tJ$-model (\ref{tj}) writes:
\begin{eqnarray}
\tilde{H}_{tJ}
& = & \tilde{\mathcal{P}}_{S}\Big{[} \sum_{\langle {\pmb i}{\pmb j} \rangle} 
         t_{{\pmb i}{\pmb j}}
      \big{[}\hat{f}^{\dagger}_{{\pmb j}}\hat{f}_{{\pmb i}}\tilde{P}_{{\pmb i}{\pmb j}}+\hc\big{]}
      \nonumber\\
&   & +\frac{1}{2}\sum_{\langle {\pmb i}{\pmb j} \rangle} J_{{\pmb i}{\pmb j}} (\tilde{P}_{{\pmb i}{\pmb j}}-1)
      \tilde{\Delta}_{{\pmb i}{\pmb j}}\Big{]}\tilde{\mathcal{P}}_{S}
      \label{angeluccihamilton}
\end{eqnarray}
where 
$\tilde{P}_{{\pmb i}{\pmb j}}=\tfrac{1}{2}(\hat{\vec{\sigma}}_{{\pmb i}}\hat{\vec{\sigma}}_{{\pmb j}}+1)$
and
$\tilde{\Delta}_{{\pmb i}{\pmb j}} =1-\hat{f}^{\dagger}_{{\pmb i}}\hat{f}_{{\pmb i}}-\hat{f}^{\dagger}_{j}\hat{f}_{{\pmb j}}$
$\tilde{\mathcal{P}}_{S} = \prod_{{\pmb i}}\big{(}1-\hat{f}^{\dagger}_{{\pmb i}}\hat{f}_{{\pmb i}}\hat{\sigma}^{-}_{{\pmb i}}
\hat{\sigma}^{+}_{{\pmb i}}\big{)}$ is a projection operator in Angelucci 
representation which projects into the subspace $S$. 
This representation (\ref{angeluccihamilton}) has two important advantages which 
facilitate numerical simulations: 
(i) Because the Hamiltonian commutes with the projection operator: 
$[\tilde{H}_{tJ} ,\tilde{\mathcal{P}}_{S}]=0$, the bare Hamiltonian ($\tilde{H}_{tJ}$ without projections) generates only states 
of subspace $S$ provided that the initial state is in the relevant subspace.
(ii) The Hamiltonian is bilinear in the spinless fermion operators.
Within the Angelucci representation the Green's function reads:
\begin{eqnarray}
G_{{\pmb j}{\pmb i}}(\tau) = \langle\hat{\sigma}^{z,+}_{{\pmb j}}(\tau) \hat{f}_{{\pmb j}}(\tau)
\hat{\sigma}^{z,+}_{{\pmb i}}(0) \hat{f}^{\dagger}_{{\pmb i}}(0) \rangle\,\, .
\end{eqnarray}
The time evolution in imaginary time is given by: 
$\hat{\sigma}^{z,+}_{{\pmb j}}(\tau) \hat{f}_{{\pmb j}}(\tau)=e^{\tau\tilde{H}_{tJ}}
\hat{\sigma}^{z,+}_{{\pmb j}}\hat{f}_{{\pmb j}} e^{-\tau\tilde{H}_{tJ}}$.
The authors of Ref. \cite{Brunner00}  show  in details how 
to implement the Green's function into the world line algorithm of our QMC simulation.  The 
spin dynamics is simulated with the loop algorithm. For each fixed 
spin configuration, one can readily compute the Green's function since the  Hamiltonian 
is bilinear  in the spinless fermion operators $\hat{f}$. 

From the  Green's function $G_{{\pmb p}}(\tau)$ we  can extract the single particle 
spectral function $A({\pmb p},\omega)$ with the  Stochastic Maximum Entropy:
\begin{eqnarray}
G_{{\pmb p}}(\tau)
=\frac{1}{\pi}\int^{\infty}_{0}d\omega e^{-\tau\omega} A({\pmb p},-\omega)\,\, .
\end{eqnarray}
In the $T=0$ limit  the asymptotic form of the 
Green's function reads:
\begin{eqnarray}
G_{{\pmb p}}(\tau) =  
|\langle\psi^{N-1}_{0}|\hat{c}_{{\pmb p}}|\psi^{N}_{0}\rangle|^{2}e^{-\mu\tau} \label{limesg}
\end{eqnarray}
where $\mu$ is the chemical potential.  As apparent, the prefactor, 
\begin{equation}
\mathcal{Z}_{\pmb p} = |\langle\psi^{N-1}_{0}|\hat{c}_{{\pmb p}}|\psi^{N}_{0}\rangle|^{2}\,\, ,
\end{equation}
is nothing but the quasiparticle residue. Hence from the asymptotic form of the 
single particle Green's function, we can read off the quasiparticle residue.

\section{Spin and Hole Dynamics}
\label{dynamics}
In this section we present our results for the spin dynamics as well as for the spectral 
function  of a doped mobile hole. 
\subsection{Spin Dynamics}
\label{spindynamics}
\begin{figure}%[h]
\centering
\includegraphics[height=4.5cm]{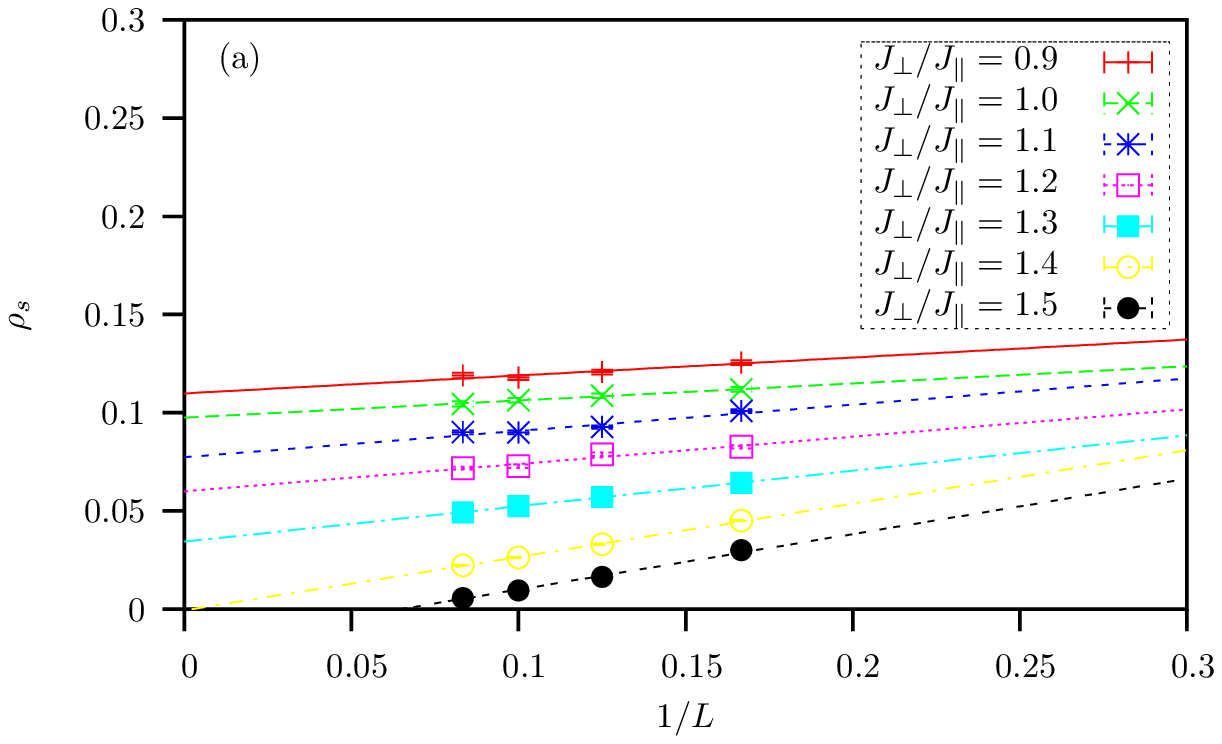} \\
\vspace{0.1cm}
\includegraphics[height=4.5cm]{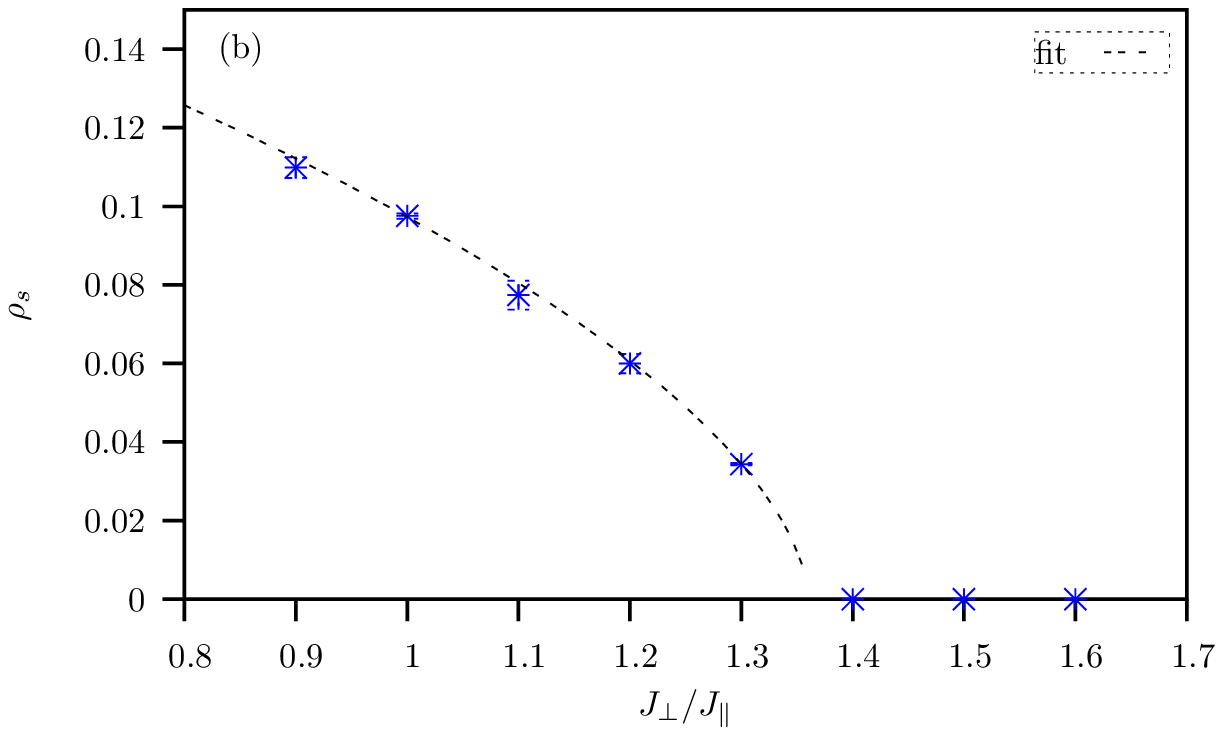}
\caption{\label{stiff}{\small (a) Spin stiffness $\rho_{s}$ as a function  
of  linear lattice size $L$  for different interplanar couplings $J_{\bot}/J_{\|}$
in the KN model. Extrapolation to the thermodynamic limit is carried out by fitting 
to the form $a + b/L$ (b) Extrapolated value of the spin stiffness  as a function of 
$J_{\bot}/J_{\|}$.  The dashes line corresponds to the fit  according to the form  of 
Eq. (\ref{Rhos_fit}).}}
\end{figure}

\begin{figure}%[h]
\centering
\includegraphics[height=4.0cm]{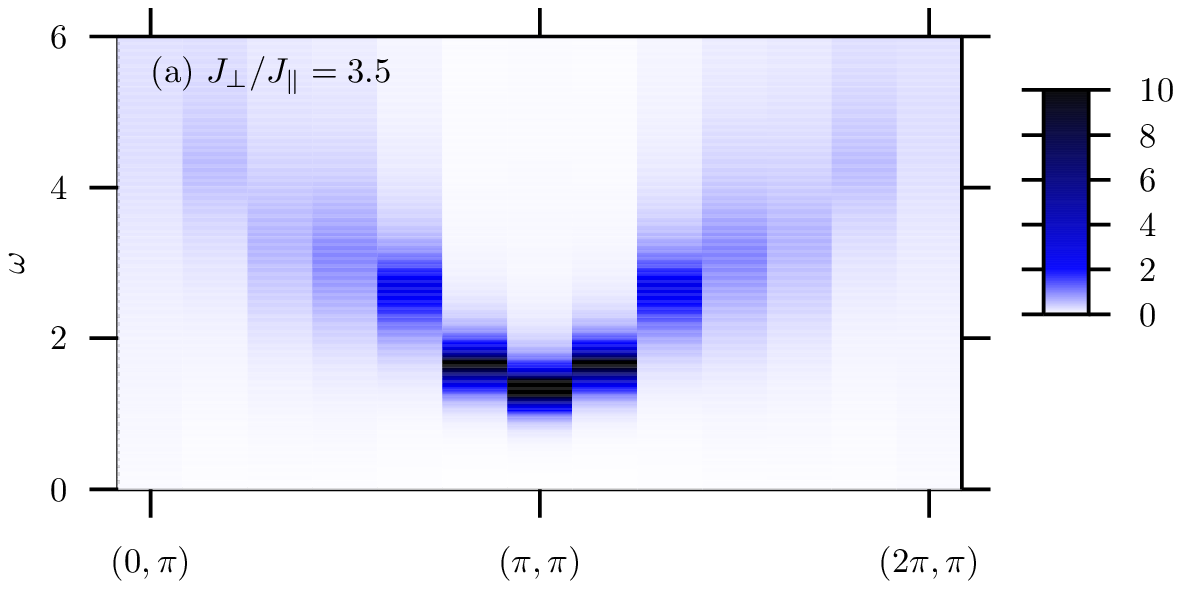}\\[0.2cm]
\includegraphics[height=4.0cm]{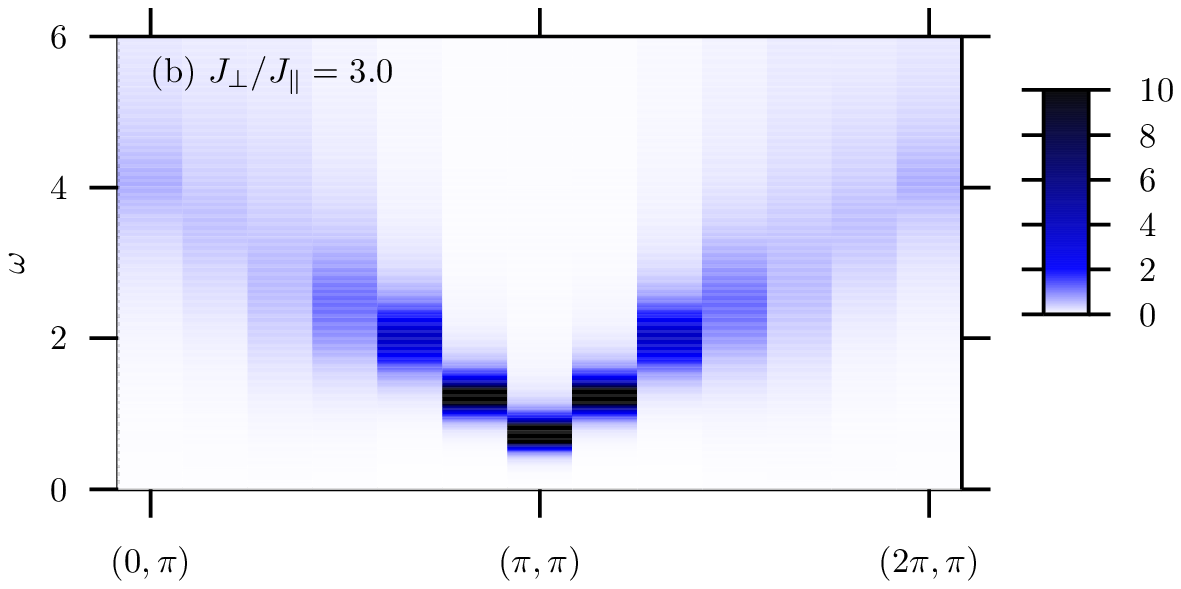}\\[0.2cm]
\includegraphics[height=4.0cm]{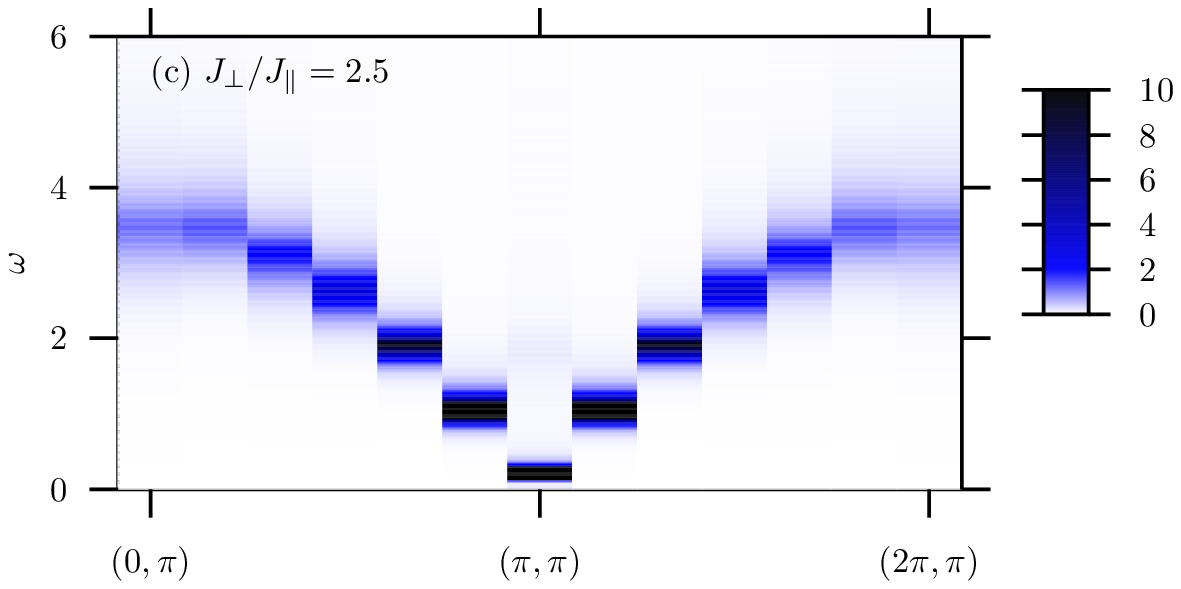}\\
${\pmb q}$

\caption{\label{magnondispersion}{\small 
Dynamical spin susceptibility, respectively magnon dispersion for different coupling ratios on
a $12\times 12$ square lattice.}}
\end{figure}

\begin{figure}%[h]
%\centering
\includegraphics[height=4.5cm]{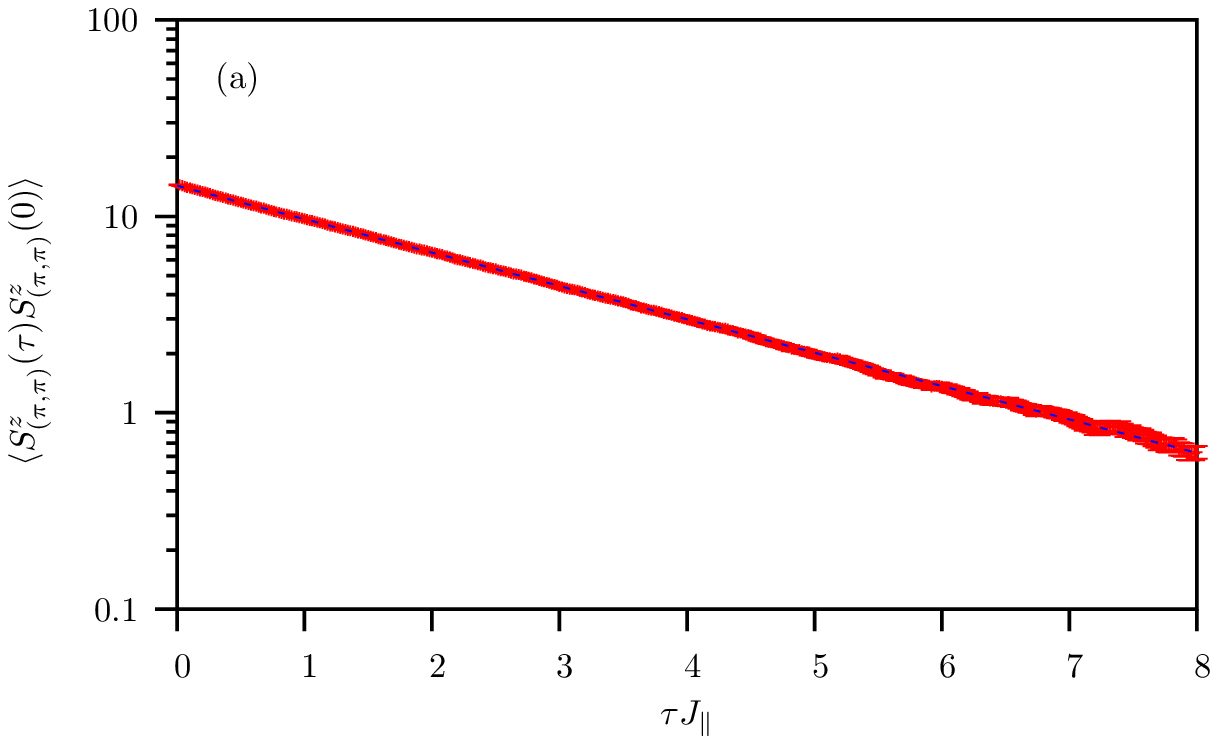}
\includegraphics[height=4.5cm]{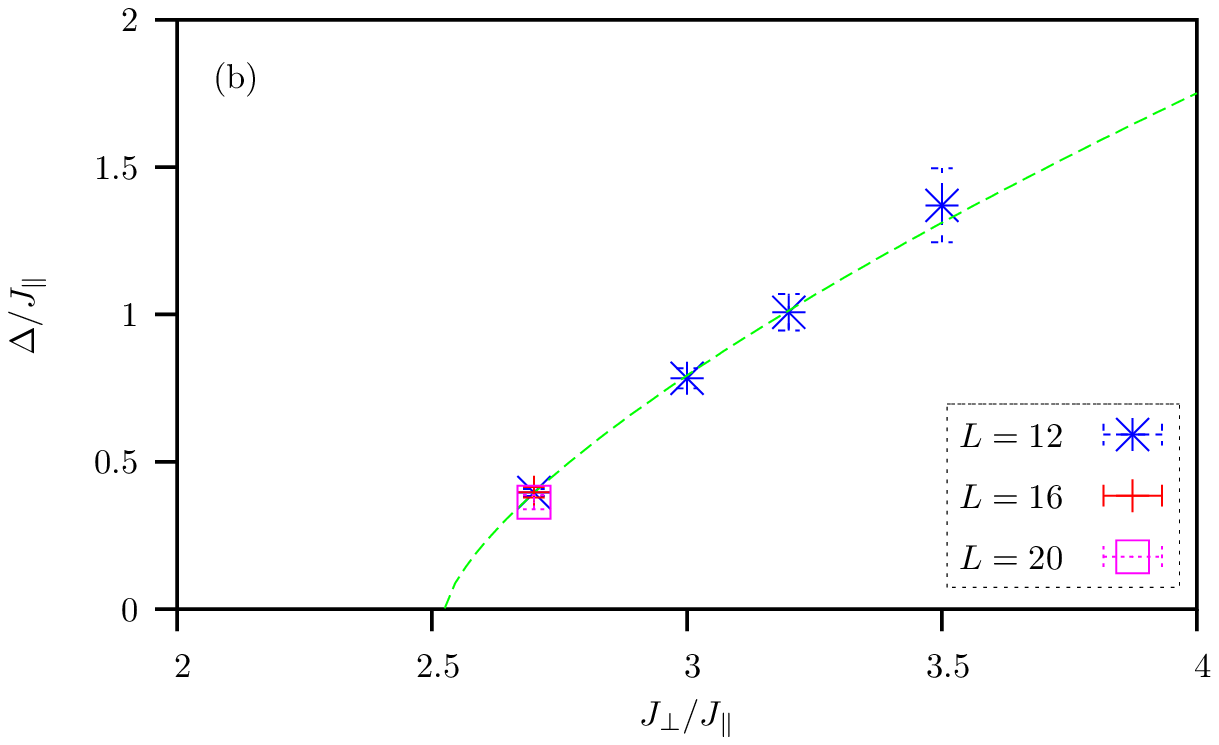}
\caption{\label{spin_gap}{\small 
(a) Spin correlation function ($J_{\bot}/J_{\|}=2.7$) 
at ${\pmb p}=(\pi,\pi)$ for a $12 \times 12$ lattice in the BHM. 
(Inverse temperature $\beta J_{\|}=30.0$, $\Delta\tau J_{\|}=0.02$)
(b) Spin gap $\Delta$ at ${\pmb p}=(\pi,\pi)$ 
for different coupling ratios $J_{\bot}/J_{\|}$. 
The data for a $12\times 12$ lattice is fitted by $\Delta\propto\left(g-g_{c}\right)^{-z\nu}$ with $g=J_{\bot}/J_{\|}$ 
and literature values $g_{c}=2.525\pm0.002$ \cite{Shevchenko00} and $z=1$ \cite{Troyer97}}}
\end{figure}

All considered models, KLM, $U$KLM, KNM and BHM, show a quantum phase 
transition between an antiferromagnetic ordered phase and a disordered phase. 
It is believed, that all models belong to the same universality class.
To demonstrate this generic 
property and to test our numerical method we determine the quantum critical point as well as
critical exponents in 
the isotropic BHM and  KNM.   Fig. 
\ref{stiff}a plots the spin stiffness  for the KNM as a function of lattice size.
The extrapolated  data is  plotted in  
Fig. \ref{stiff}b. We fit the data to the form:
\begin{eqnarray}
\rho_s \propto\Big{[}\Big{(}\frac{J_{\bot}}{J_{\|}}\Big{)}_{c}-\Big{(}\frac{J_{\bot}}{J_{\|}}
\Big{)}\Big{]}^{\nu}\,\, 
\label{Rhos_fit}
\end{eqnarray} 
to obtain $(J_{\bot}/J_{\|})_{c}=1.360\pm 0.017$ and
a critical exponent of $\nu =0.582\pm 0.077$, which agrees 
(within the error bars) with the value of Ref. 
\cite{Troyer97}: $\nu=0.685\pm 0.035$.
Similar data for the  BHM localizes the quantum critical point at 
$(J_{\bot}/J_{\|})_{c}=2.5121\pm 0.0044$, which conforms roughly the literature value 
$(J_{\bot}/J_{\|})^{lit}_{c}=2.525\pm 0.002$ of 
 Ref. \cite{Shevchenko00}. For the critical exponent we obtain 
$\nu=0.7357\pm 0.044$. Again this is  in good agreement with the critical exponent specified
in Refs. \cite{Troyer97}. In Ref. \cite{Kotov98} the BHM and the KNM are observed by 
dimer series expansions. Within this framework our numerical results are reflected quite well.
FIG. \ref{magnondispersion} plot the dynamical spin structure factor as a function
of  $J_{\bot}/J_{\|}$ for the BHM. 
In the deeply disordered phase the dispersion has a cosine-like shape. 
In the limit 
$J_{\bot}\to \infty$ the ground state wave function is a tensor product of singlets in 
each unit cell.  Starting from this state,  a magnon corresponds to  breaking a singlet 
to form a triplet. In first order perturbation theory in $J_{\bot}/J_{\|}$, the magnon 
acquires a dispersion relation:
\begin{eqnarray}
\Omega ({\pmb q})\approx J_{\bot}+\tfrac{1}{2}J_{\|}\gamma ({\pmb q})
\label{mdispstrong}
\end{eqnarray}
with $\gamma({\pmb q})=2\left(\cos(q_{x})+\cos(q_{y})\right)$. This approximative approach 
is roughly consistent with the large-$J_{\bot}$ case in Fig. \ref{magnondispersion}a. 
As as  function of decreasing coupling $J_{\bot}$ the spin gap progressively closes (see 
 Fig. \ref{spin_gap}) and at the critical coupling the magnons at ${\pmb q}=(\pi,\pi)$ 
condense to form the antiferromagnetic order. This physics is captured by the bond mean field 
approximation which we discuss below. 
\\[0.2cm]

\subsubsection*{Bond Operator Mean Field Approach}
The bond mean field approach \cite{Sachdev90} is a strong coupling approximation in $J_{\bot}$. 
The spins between layers dominantly form singlets and the density of triplets is 
"low". This assumption allows one to neglect triplet-triplet interaction. 
The bond operator representation describes the system
in a base of pairs of coupled spins, which can either be in a singlet
or triplet state.
\begin{eqnarray}
|s\rangle_{\vi}
&=& \hat{s}^{\dagger}_{\vi}|0\rangle_{\vi}=
    \frac{1}{\sqrt{2}}(|\uparrow\downarrow\rangle_{\vi}-|\downarrow\uparrow\rangle_{\vi}) \label{bonds}
    \nonumber\\
|t_{x}\rangle_{\vi}
&=& \hat{t}^{\dagger}_{\vi x}|0\rangle_{\vi}=
    \frac{-1}{\sqrt{2}}(|\uparrow\uparrow\rangle_{\vi}-|\downarrow\downarrow\rangle_{\vi})
    \nonumber\\
|t_{y}\rangle_{\vi}
&=& \hat{t}^{\dagger}_{\vi y}|0\rangle_{\vi}=
    \frac{\ii}{\sqrt{2}}(|\uparrow\uparrow\rangle_{\vi}+|\downarrow\downarrow\rangle_{\vi})
    \nonumber\\
|t_{z}\rangle_{\vi}
&=& \hat{t}^{\dagger}_{\vi z}|0\rangle_{\vi}=
    \frac{1}{\sqrt{2}}(|\uparrow\downarrow\rangle_{\vi}+|\downarrow\uparrow\rangle_{\vi}) \label{bondtz}
\end{eqnarray}
The operators $\hat{t}^{\dagger} $ and $\hat{s}^{\dagger}$ satisfy Bose commutation rules 
provided that we impose the constraint 
\begin{eqnarray}
\sid\si+\sum_{\alpha}\tialphad\tialpha = 1 \,\, .
\label{constraint}
\end{eqnarray}
Since the original spin 1/2 degrees of freedom reads, 
\begin{eqnarray}
\hat{S}^{(1,2)}_{\vi \alpha}
& = & \tfrac{1}{2}(\pm\sid\tialpha\pm\tialphad\si-\ii\sum_{\beta\gamma}
\epsilon_{\alpha\beta\gamma}\tibetad\tigamma)\,\, ,
\end{eqnarray}
the Hamiltonian (\ref{starth}) can  be rewritten in the bond operator representation as:
\begin{eqnarray}
\tilde{H}
& = & J_{\bot}\sum_{{\pmb i}}\big{(}
      -\tfrac{3}{4}\sid\si+\tfrac{1}{4}\sum_{\alpha}\tialphad\tialpha\big{)}
      \nonumber\\
&   &  -\sum_{{\pmb i}}\mu_{{\pmb i}}\big{(}\sid\si+\sum_{\alpha}\tialphad\tialpha-1\big{)}
      \nonumber\\
&   & +\frac{J_{\|}}{2}\sum_{\langle{\pmb i}{\pmb j}\rangle}\sum_{\alpha}
      \big{(}\sid\sjd\tialpha\tjalpha+\sid\sj\tialpha\tjalphad+\hc\big{)}
      \nonumber\\
&   & +\frac{J_{\|}}{2}\sum_{\alpha,\beta,\gamma}(\tibetad\tigamma\tjbetad\tjgamma-
      \tibetad\tigamma\tjgammad\tjbeta)\,\, .
      \nonumber
\end{eqnarray}
$\mu_{{\pmb i}}$ is a Lagrange parameter which enforces locally the constraint (\ref{constraint}).
The interplanar part shows the characteristic Hamiltonian of two antiferromagnetically 
coupled spins whereas the intraplanar part includes the interaction between singlets and
triplets of different bonds. 
We now follow the standard method of Sachdev and Bhatt \cite{Sachdev90}.
In the disordered phase we expect a singlet
condensate ($\bar{s}=\langle s\rangle\ne 0$) and impose the constraint only on average 
($\mu_{\pmb i}=\mu$). As mentioned above we neglect triplet-triplet
interactions. 
Apart from a constant we obtain the following mean field 
Hamiltonian in momentum space:
\begin{eqnarray}
\hat{H}_{MFA}
& = & \sum_{\alpha}\sum_{{\pmb q}}A_{\pmb q}
      \hat{t}^{\dagger}_{{\pmb q}\alpha}\hat{t}_{{\pmb q}\alpha}
      \nonumber\\
&   &  +\sum_{\alpha}\sum_{{\pmb q}}\frac{B_{\pmb q}}{2}
      (\hat{t}^{\dagger}_{{\pmb q}\alpha}\hat{t}^{\dagger}_{-{\pmb q}\alpha}+\hc)
      \label{mfh}\,\, ,
\end{eqnarray} 
where
\begin{eqnarray}
A_{\pmb q}
& = & \frac{J_{\bot}}{4}-\mu+J_{\|}\bar{s}^{2}
      \big{(}\cos(q_{x})+\cos(q_{y})\big{)}
\\
B_{\pmb q}
& = & J_{\|}\bar{s}^2
      \big{(}\cos(q_{x})+\cos(q_{y})\big{)}\,\, .
\end{eqnarray} 

The parameter $\mu$ and $\bar{s}=\langle s\rangle$ are
determined by the saddle-point equations: 
$\langle\partial\hat{H}_{MFA}/\partial\mu\rangle=0$ 
and 
$\langle\partial\hat{H}_{MFA}/\partial\bar{s}\rangle=0$.
The Hamiltonian is diagonalized by a Bogoliubov transformation: 
$\hat{\alpha}^{\dagger}_{{\pmb q}\alpha}
 =  u_{{\pmb q}}\hat{t}^{\dagger}_{{\pmb q}\alpha}-
      v_{{\pmb q}}\hat{t}_{-{\pmb q}\alpha}$. 
In terms of magnon creation and annihilation operators the Mean field Hamiltonian 
(\ref{mfh}) writes:
\begin{eqnarray}
\hat{H}_{MFA} = \sum_{\pmb q}\sum_{\alpha}\Omega({\pmb q})
                \hat{\alpha}^{\dagger}_{{\pmb q}\alpha}\hat{\alpha}_{{\pmb
                q}\alpha}\,\, .
                \label{diagham}
\end{eqnarray}
The Bogoliubov coefficients $u_{{\pmb q}}$ and $v_{{\pmb q}}$ 
satisfy the relation $u^{2}_{{\pmb q}}-v^{2}_{{\pmb q}}=1$, which follows 
from the bosonic nature of the magnons: 
$[\hat{\alpha}_{{\pmb q}},\hat{\alpha}^{\dagger}_{{\pmb q}'}]=\delta_{{\pmb q}{\pmb q}'}$. 
The coefficients are given by
\begin{eqnarray}
u_{{\pmb q}},v_{{\pmb q}}
&=& \sqrt{\frac{A_{\pmb q}}{2\Omega({\pmb q})}\pm\frac{1}{2}}\,\, ,
\label{uv}
\end{eqnarray}
where 
$\Omega({\pmb q})
= \sqrt{A^{2}_{\pmb q}-B^{2}_{\pmb q}}$
is the magnon dispersion. In the vicinity of the critical point it can be approximated by
\begin{eqnarray}
\Omega({\pmb q})=\sqrt{\Delta^{2}+v^{2}_{s}({\pmb q}-{\pmb Q})^{2}}
\label{mdisp}
\end{eqnarray}
with $\Delta$ the energy gap to magnon excitations, $v_{s}$ the magnon
velocity and ${\pmb Q}=(\pi,\pi)$. Eq. (\ref{mdisp}) gives an accurate description 
of the dispersion relation in the vicinity of 
the critical point (see FIG. \ref{magnondispersion}c).
At the critical point the gap $\Delta$ vanishes, 
so that the triplets can condense thus forming the AF static ordering.

\subsection{Hole dynamics}
\label{holedynamics}
We now dope our systems with a single mobile hole and  restrict its  motion 
to one layer thereby staying in the spirit of Kondo lattice models.   
To understand the coupling of the hole to magnetic 
fluctuations within the magnetic disordered  phase we can extend the  previously described 
bond mean-field approximation (See Eq. (\ref{diagham})) to account for the hole motion. 
For this 
we introduce the operator $\hat{h}^{\dagger}_{{\pmb i}\sigma}$  ($\hat{h}_{{\pmb i}\sigma}$), 
that 
creates (anhilates) a hole with spin $\sigma$ in layer 1 at site ${\pmb i}$. 
\begin{eqnarray}
\hat{h}^{\dagger}_{{\pmb i}\sigma}|vac\rangle&=&|0\sigma\rangle_{{\pmb i}}
\end{eqnarray}
$|\sigma_{1}\sigma_{2}\rangle_{\pmb i}$ denotes a dimer state at site ${\pmb i}$ 
with spin $\sigma_{1}$ in layer 1 and spin $\sigma_{2}$ in layer 2.
The Hamiltonian now writes \cite{Vojta99}:
\begin{eqnarray}
\hat{H}&=&\sum_{{\pmb q}}\Omega({\pmb q})
          \hat{{\pmb \alpha}}^{\dagger}_{{\pmb q}}\hat{{\pmb \alpha}}_{{\pmb q}}
          +\sum_{{\pmb p}}\varepsilon({\pmb p})
          \hat{h}^{\dagger}_{{\pmb p}} \hat{h}_{{\pmb p}}
          \label{holemagnonhamilton}\\
& &       +\sum_{{\pmb p},{\pmb q}}g({\pmb p},{\pmb q})\,{\pmb \alpha}_{\pmb q}\cdot
          \big{(}\hat{h}^{\dagger}_{{\pmb p}+{\pmb q}} {\pmb \sigma} \hat{h}_{{\pmb p}}\big{)} + \text{h.c.}
          \nonumber
\end{eqnarray}
with spinor $\hat{h}_{{\pmb p}}=(\hat{h}_{{\pmb p}\uparrow},\hat{h}_{{\pmb
p}\downarrow})$ and vector 
$\hat{{\pmb \alpha}}_{{\pmb q}}=(\hat{\alpha}_{{\pmb q}x},
\hat{\alpha}_{{\pmb q}y},\hat{\alpha}_{{\pmb q}z})$.
${\pmb \sigma}=(\sigma^{1},\sigma^{2},\sigma^{3})$ denotes the Pauli matrices. 
The coupling strength between the hole and magnons is given by $g({\pmb p},{\pmb q})$. 
We discuss $g({\pmb p},{\pmb q})$ in detail later in section \ref{qpr}.
For the bare hole dispersion the calculation yields
\begin{eqnarray}
\varepsilon({\pmb p})=+t\bar{s}^{2}\big{(}\cos (p_{x})+\cos (p_{y})\big{)}\,\, .
\end{eqnarray}

In the limit $J_{\bot}\to\infty$ the magnon excitation energy diverges 
(see Eq. (\ref{mdispstrong})) and hence the coupling of the hole to magnetic excitations 
becomes negligible.
In this limit the magnon excitations become quite rare, so that:  
$\bar{s}\equiv\langle s\rangle\approx 1$. 
Thus, in the strong coupling region we obtain from (\ref{holemagnonhamilton})  
a hole dispersion relation:
\begin{eqnarray}
E({\pmb p})=t\big{(}\cos (p_{x})+\cos (p_{y})\big{)}\,\, .
\label{strongcoupling}
\end{eqnarray}   
This agrees with the result given by applying perturbation theory in $t/J_{\bot}$ 
\cite{Tsunetsugu97}.

\begin{figure}%[h]
%\centering
(a) $J_{\bot}/J_{\|}=10.0$\\[0.3cm]
\includegraphics[height=4.5cm]{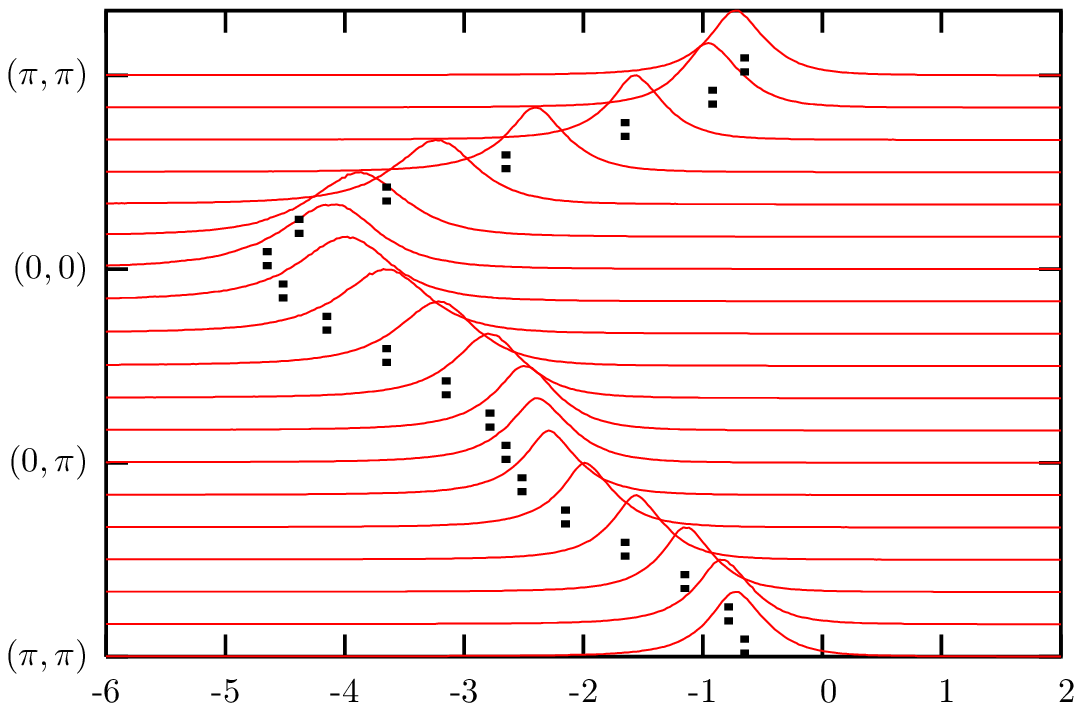}\\[0.0cm]
(b) $J_{\bot}/J_{\|}=2.4$\\[0.3cm]
\includegraphics[height=4.5cm]{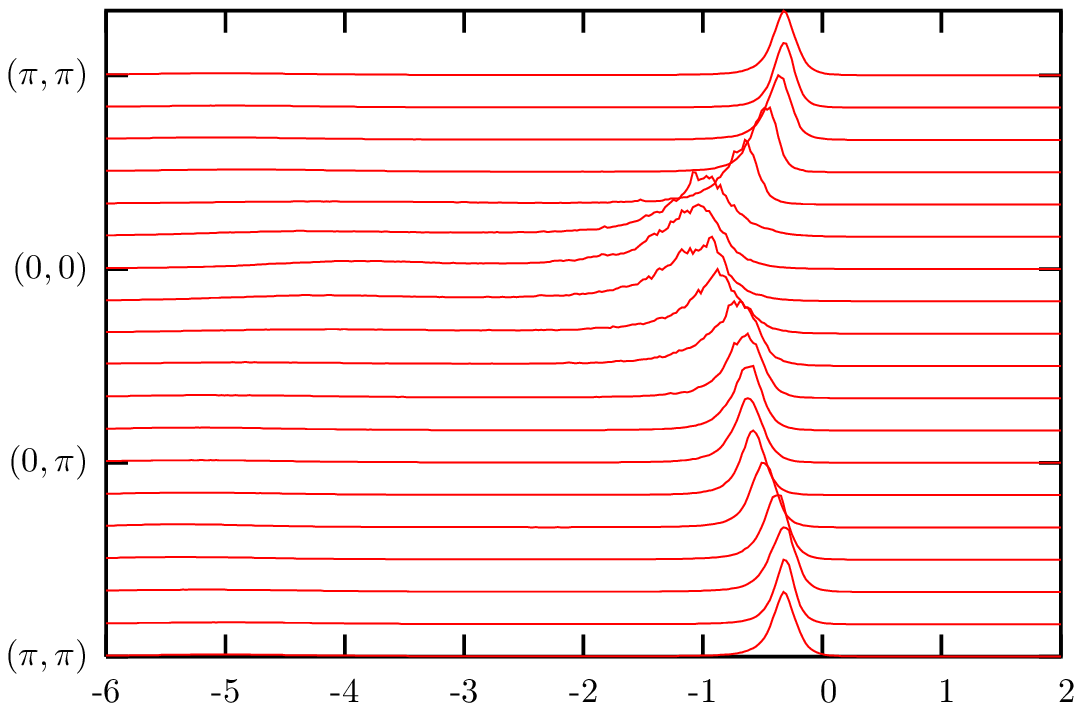}\\[0.0cm]
(c) $J_{\bot}/J_{\|}=2.0$\\[0.3cm]
\includegraphics[height=4.5cm]{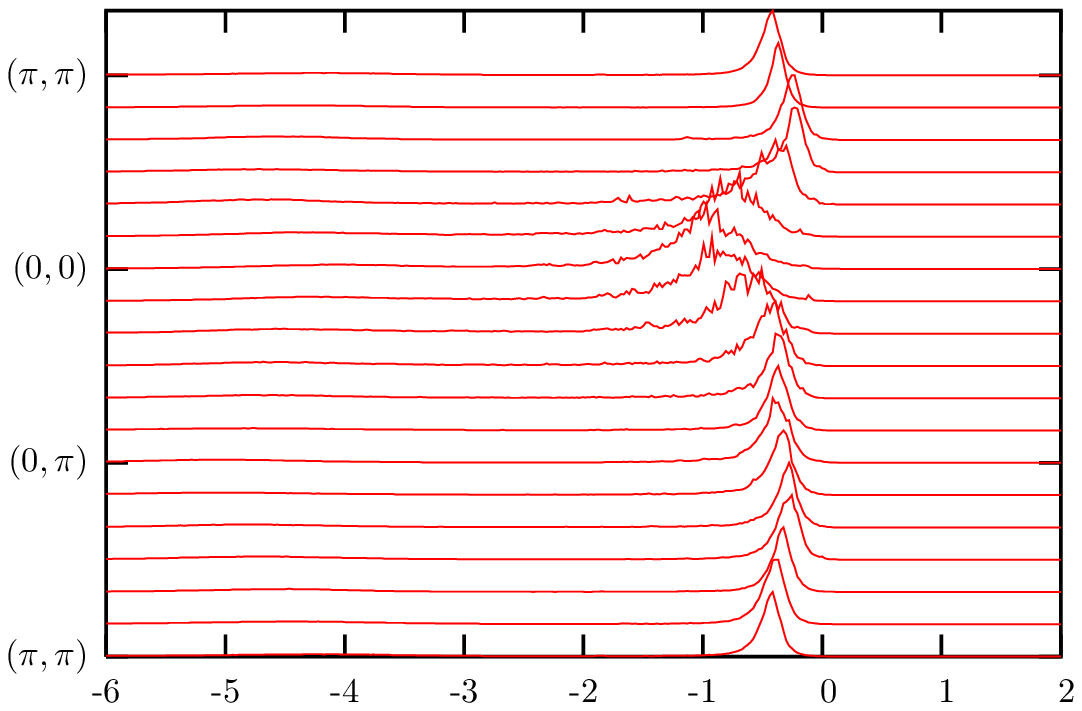}\\[0.0cm]
(d) $J_{\bot}/J_{\|}=1.0$\\[0.3cm]
\includegraphics[height=4.5cm]{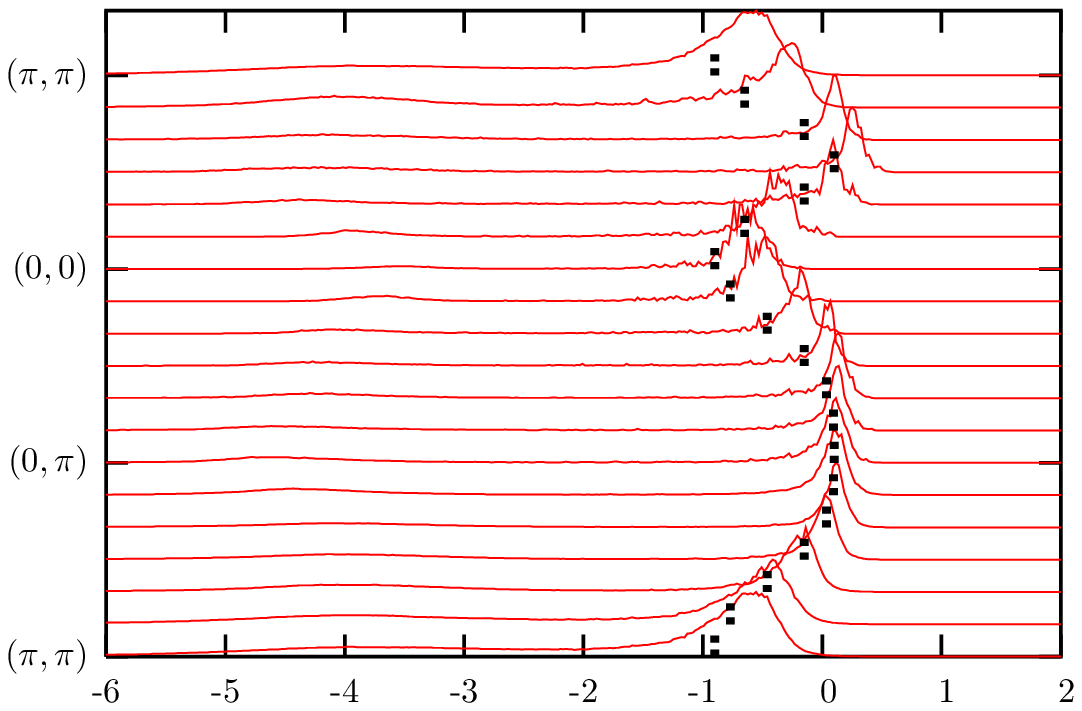}
$\omega/J_{\|}$
\caption{\label{dynamic_spec}{\small Spectrums of a mobile hole for a $12\times 12$ lattice 
in the BHM. The small dashed lines in (a) tag the dispersion of a free particle; in (d) they 
outline a dispersion of the form: 
$E ({\pmb p})=J_{\|}\left( \cos(p)_{x} + \cos(p_{y})\right)^{2}$.}}
\end{figure}

\begin{figure}%[h]
%\centering
(a) $J_{\bot}/J_{\|}=4.0$\\[0.3cm]
\includegraphics[height=4.5cm]{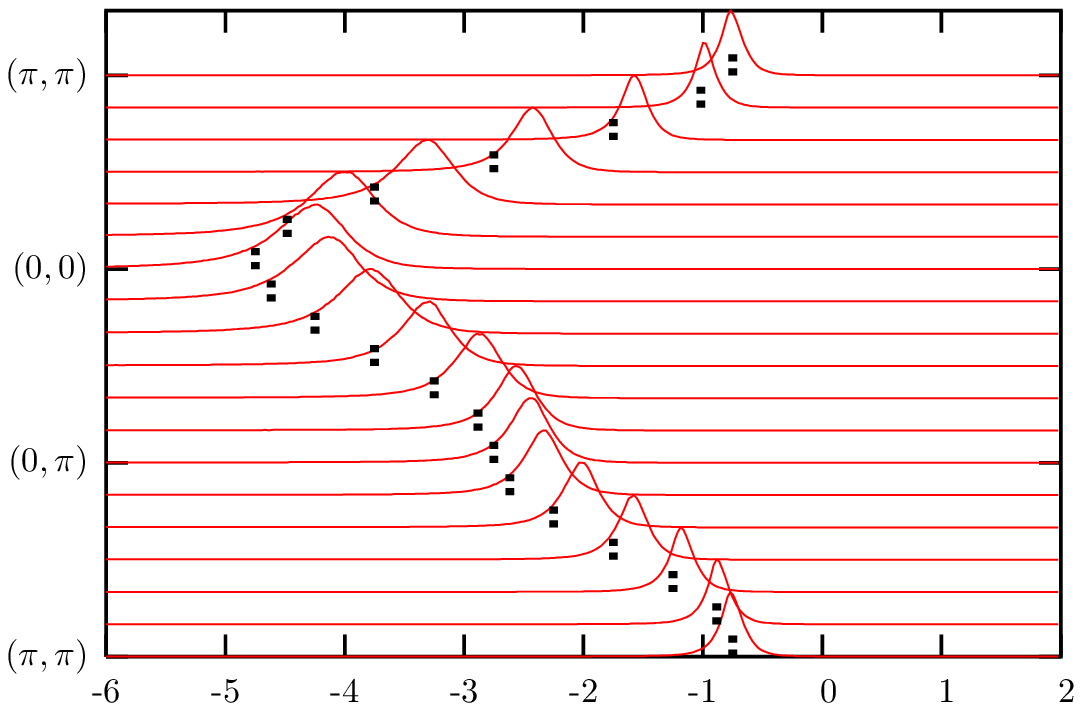}\\[0.0cm]
(b) $J_{\bot}/J_{\|}=2.0$\\[0.3cm]
\includegraphics[height=4.5cm]{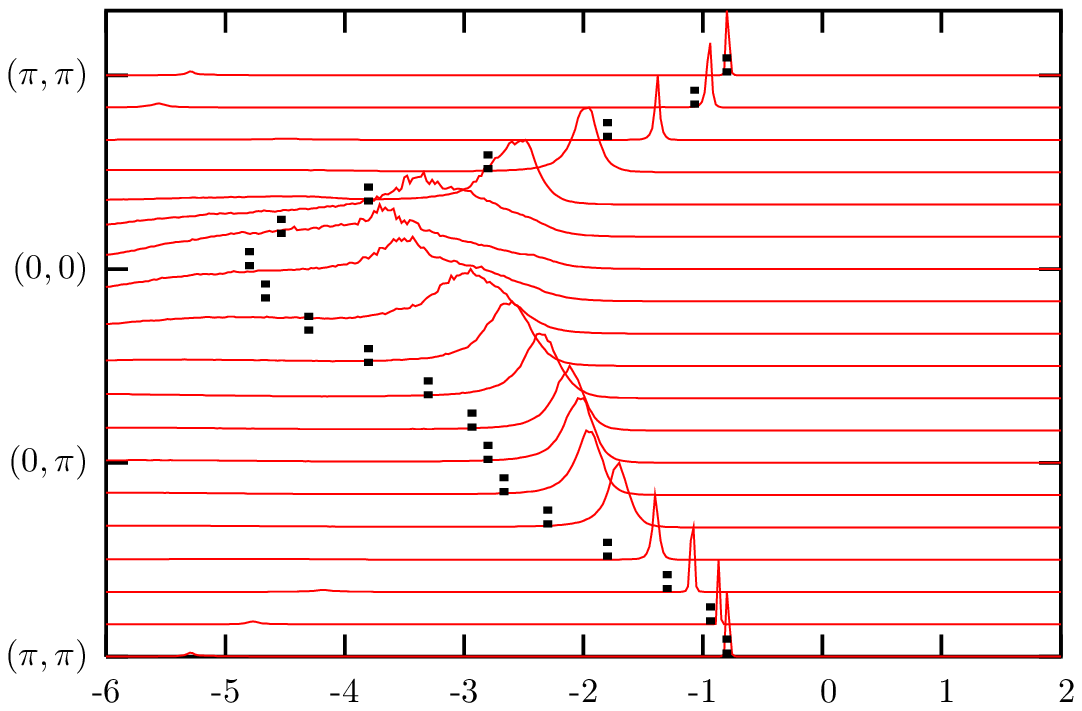}\\[0.0cm]
(c) $J_{\bot}/J_{\|}=0.5$\\[0.3cm]
\includegraphics[height=4.5cm]{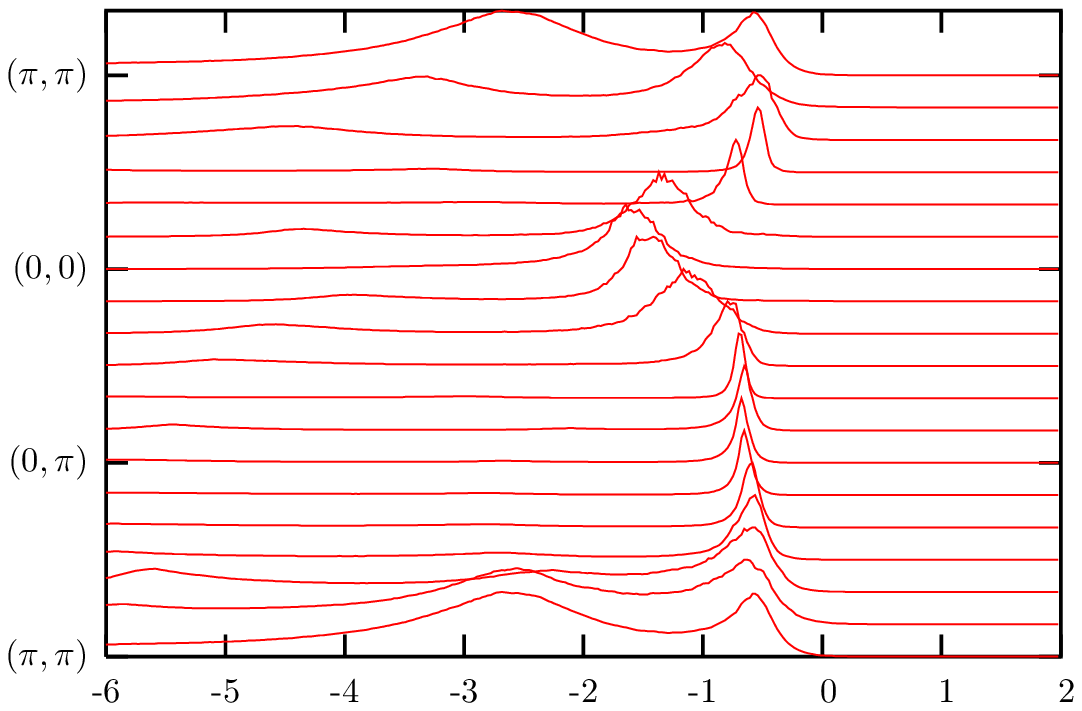}
$\omega/J_{\|}$
\caption{\label{kondo_spec}{\small Spectrum of the KNM for a $12\times 12$ lattice. The dashed lines 
tag the dispersion of a free particle.}}
\end{figure}

As apparent from Figs. \ref{dynamic_spec} and  \ref{kondo_spec}   this strong  coupling 
behavior is reproduced by the Monte Carlo simulations where 
the dispersion  exhibits a cosine form with maximum at ${\pmb p}=(\pi,\pi)$.   The form of 
this dispersion relation directly reflects the singlet formation --  in other words Kondo 
screening --   between spin degrees of freedom on different layers.   We note that this strong 
coupling behavior  of the dispersion relation sets in at larger values of $J_{\bot}/J_{\|}$ for 
the BHM than for the KNM. 
This is quite reasonable since in the BHM the single bonds are coupled among each other 
within both layers.

With decreasing coupling ratio  the bandwidth of the quasiparticle dispersion relation 
diminishes  but the overall features of the strong coupling remain. 

In the weak coupling limit we observe considerable differences between the single particle 
spectrum of the BHM and KNM.  Let us start with the BHM.  For this model the point 
$J_{\bot}/J_{\|} =0$ is well defined (i.e. the ground is  non-degenerate on 
any finite lattice) and corresponds to two independent Heisenberg planes 
with  mobile hole in the upper plane. The problem of the single hole in a two dimensional 
Heisenberg model has been addressed in the framework of the self-consistent  Born approximation 
\cite{Martinez91}, and yields a dispersion relation  given by:
\begin{eqnarray}
\label{SCB_Horsch}
E ({\pmb p})=J_{\|}\left( \cos(p)_{x} + \cos(p_{y})\right)^{2}\,\, .
\end{eqnarray}
Since at $J_{\bot}/J_{\|} =0$  we have a well defined ground state  we can expect that turning 
on  a small value of $J_{\bot}/J_{\|}$  will not alter the
single hole dispersion relation. This point of view is confirmed in Fig. \ref{dynamic_spec}. At 
$J_{\bot}/J_{\|} = 1$, the single hole dispersion relation follows of Eq. (\ref{SCB_Horsch}).
Hence and  as confirmed by Fig. \ref{dynamic_spec} the dispersion relation of a single hole 
in the BHM continuously deforms from the strong coupling form of Eq. (\ref{strongcoupling}) to 
that of a doped hole in a planar antiferromagnet (see Eq. (\ref{SCB_Horsch})). Hence  as 
a function of $J_{\bot}/J_{\|}$ there is a point where the effective mass  (as defined by 
the inverse curvature of the dispersion relation) at $ {\pmb p} = (\pi,\pi) $ diverges.  Upon 
inspection of the data (see Fig. \ref{dynamic_spec}), the point of divergence of the effective
mass is not related to the magnetic quantum phase transition and  since it occurs slightly 
below $\left( J_{\bot}/J_{\|}\right)_c$. This crossover between a dispersion with minimum 
at $ {\pmb p} = (\pi,\pi) $ and minimum at $ {\pmb p} = (\pi/2,\pi/2) $ with a crossover point 
lying inside the AF ordered phase is also documented in Ref. \cite{Vojta99}.

The above argument can not be  applied to the KNM, since the $J_{\bot}/J_{\|} =0$ point 
is macroscopically degenerate and hence is not a  good starting point to understand the 
weak-coupling physics.   Clearly the same holds for the KLM and UKLM.
Inspection of the spectral data  deep in the ordered phase of the KNM 
(see Fig. \ref{kondo_spec}c)  shows that  the maximum of the dispersion relation is still 
pinned at ${\pmb p}=(\pi,\pi) $  such that the strong coupling features  stemming from 
Kondo screening is still present at weak couplings.  For the KNM and up to the lowest  
couplings we have considered the effective mass at ${\pmb p} = (\pi,\pi) $ increases as 
a function of decreasing coupling strength  but does not seem to diverge  at finite 
values of $J_{\bot}/J_{\|} $.  Precisely the same conclusion is reached in the 
framework of the KLM \cite{Capponi00} and UKLM \cite{Feldbacher02}.

\section{Quasi Particle Residue}
\label{qpr}
In this section we  turn out attention to the delicate issue of the 
quasiparticle residue in the vicinity of the magnetic quantum phase transition.  
We first address this question within the framework of the the mean-field model 
of Eq. (\ref{holemagnonhamilton})  and compute the single particle  Green's function 
within the framework of the self-consistent Born approximation.  In a second step, 
we attempt to determine the quasiparticle residue directly from the Monte Carlo data. 
\subsection{Analytical Approach}
\label{Analytical.sec}
\begin{figure}%[h]
\centering
\includegraphics[width=0.4\textwidth]{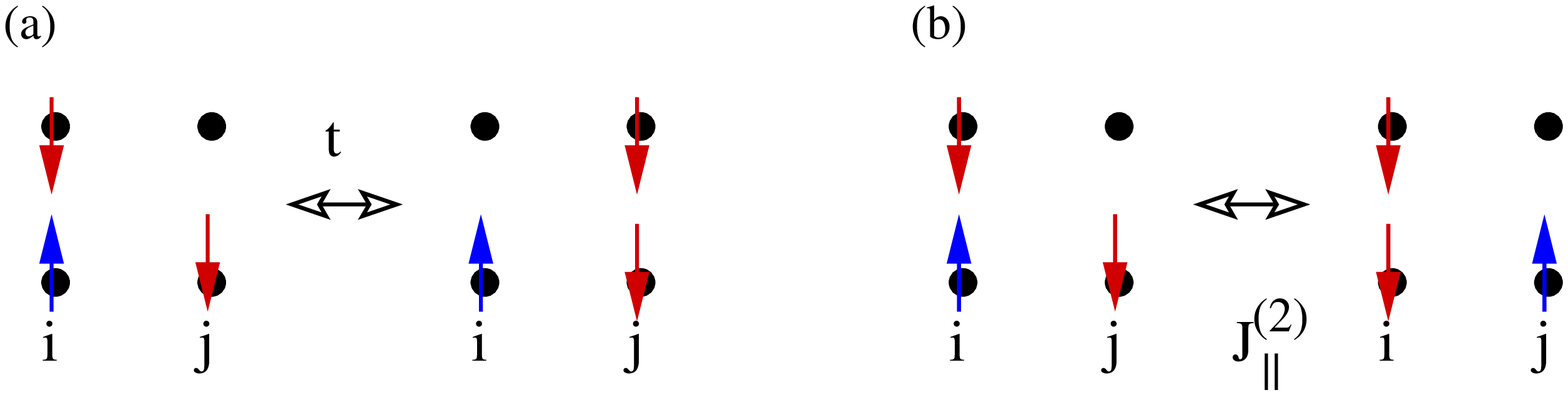}
\caption{\label{coupling}{\small Two possible processes where the hole can couple to
 magnons: (a) The hole moves to a next neighbor. (b) The hole is at rest.}}
\end{figure}

Here we restrict our analysis to the BHM. 
and return to the Hamiltonian (\ref{holemagnonhamilton}).  The coupling between the 
hole and magnons  $g({\pmb p},{\pmb q})$  reads:
\begin{eqnarray*}
      \quad g({\pmb p},{\pmb q})=g_{a}({\pmb p},{\pmb q})+g_{b}({\pmb p},{\pmb q}).
\end{eqnarray*}
We identify the two coupling constants with the processes that are shown in
Fig. \ref{coupling}: $g_{a}({\pmb p},{\pmb q})$ is proportional to the hopping matrix element and 
hence describes the coupling of a mobile hole to magnetic background, whereas  
$g_{b}({\pmb p},{\pmb q})$   is proportional to $J^{(2)}_{\|}$ and describes  the coupling of a 
hole at rest with the magnons.
Our calculations give the following momentum dependent coupling strengths:
\begin{eqnarray}
g_{a}({\pmb p},{\pmb q})&=&
-\frac{t\bar{s}}{\sqrt{N}}\big{(}\gamma({\pmb p}+{\pmb q})u({\pmb q})+\gamma({\pmb p})v({\pmb q})\big{)}\\
g_{b}({\pmb p},{\pmb q})&=&
-\frac{J^{(2)}_{\|} \bar{s}}{8\sqrt{N}}\gamma({\pmb q})\big{(}u({\pmb q})+v({\pmb q})\big{)}
\end{eqnarray}
where $\gamma({\pmb q})=2\big{(}\cos (q_{x})+\cos (q_{y})\big{)}$.
We concentrate on the coupling to critical magnetic fluctuations and hence set 
${\pmb q}={\pmb Q}$  and place  ourselves in the proximity of the quantum phase transition, 
on the disordered side. 
In this case  $ \Omega( {\pmb Q} ) \rightarrow 0 $ and the coherence factors 
(see Eq. (\ref{uv})) are both proportional to $ \Omega({\pmb q})^{-\frac{1}{2}} $. 
Since furthermore  $\gamma({\pmb p}+{\pmb Q})=-\gamma({\pmb p})$   one arrives at the conclusion
that $g_{a}({\pmb p},{\pmb Q})$ vanishes at the critical point. 
There is hence no coupling via process (a) to critical fluctuations.  In other words 
process (a) couples only to short range spin  fluctuations.   
On the other hand  in the vicinity of the  critical point  $g_{b}$ scales  as 
$g_{b}({\pmb p},{\pmb q}) \propto  \Omega({\pmb q})^{-\frac{1}{2}} $ so that we  can only 
retain this term to understand the coupling to critical fluctuations. Summarizing we set:
\begin{eqnarray}
g({\pmb p},{\pmb q}) \to g_{b} (\pmb q) \propto\frac{1}{\sqrt{\Omega(\pmb q)}}\,\, ,
\end{eqnarray}
for the subsequent calculations.   It is intriguing to note that in this simple approximation
$ g_{b} (\pmb q) $ scales as $J^{(2)}_{\|} $, which is strictly speaking  null in the KNM. 
However, such a coupling should be dynamically generated via an RKKY-type interaction. 

\begin{figure}%[h]
%\centering
\includegraphics[height=4.5cm]{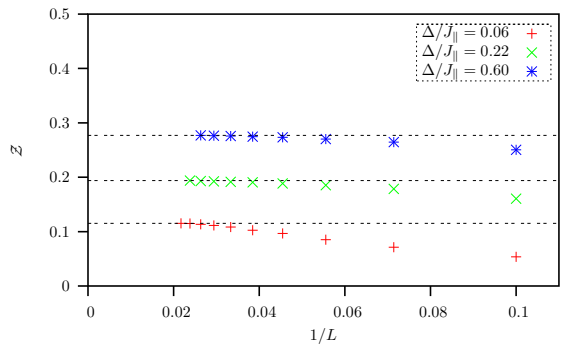}
\caption { \label{scb1}  Self consistent Born approximation: QPR as a function of 
linear  lattice size, 
$L$, for different spin gap energies, $\Delta$. } 
\end{figure}

\begin{figure}%[h]
\includegraphics[height=4.5cm]{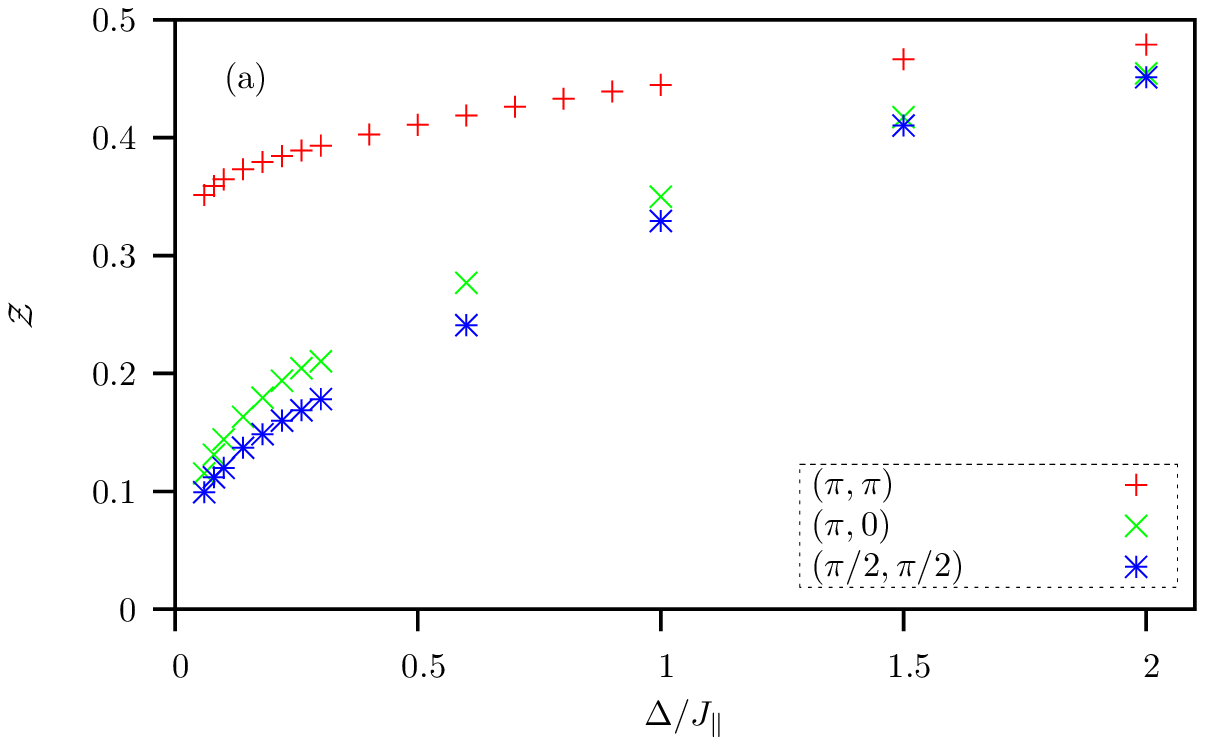}\\
\includegraphics[height=4.5cm]{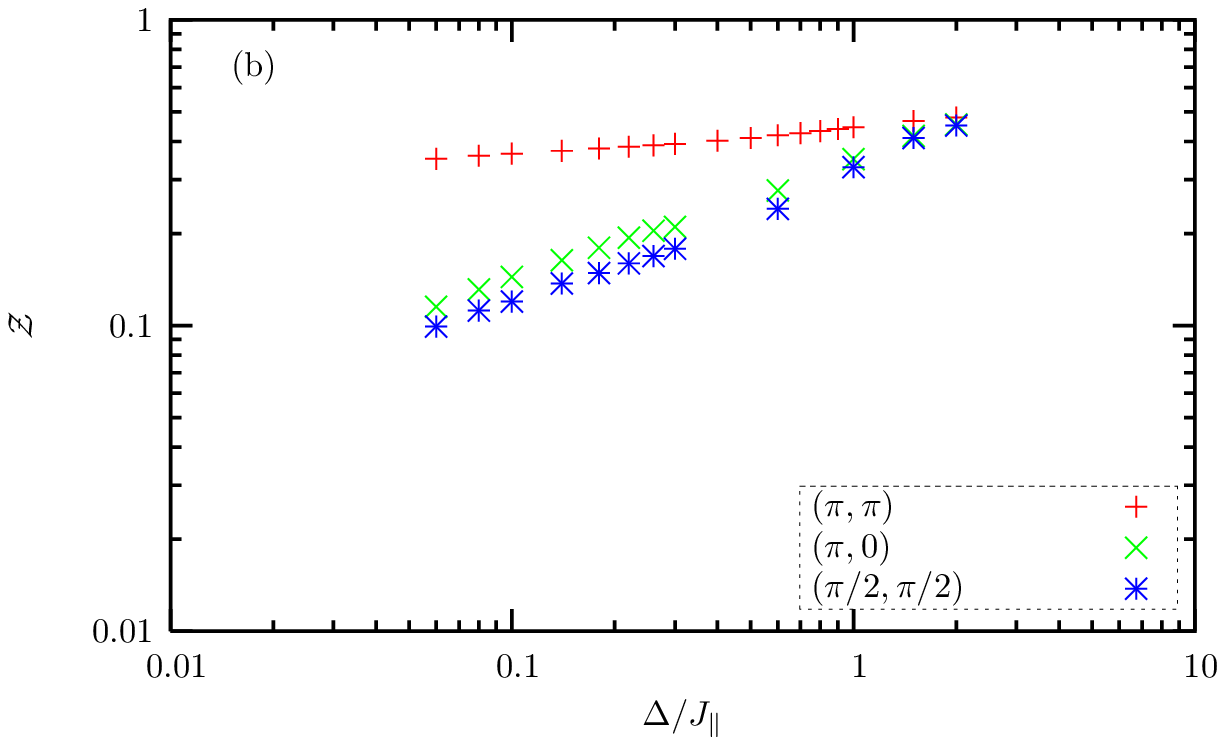}
\caption{\label{scb2}  Self consistent Born approximation: QPR in the vicinity of the 
quantum critical point for selected hole momenta (a) in a linear plot and (b) in a double logarithmic plot. 
$\Delta$ corresponds to the spin gap.}
\end{figure}

With the above couple the  first order self energy diagram for wave vectors satisfying  
$ \epsilon({\pmb p}) = \epsilon({\pmb p} + {\pmb Q} ) $  shows a logarithmic divergence 
as a function of the spin gap.  Hence we have to sum up all diagrams. We do so in 
the non-crossing or self-consistent Born approximation which in the $T=0$ limit boils down
to the following set of self-consistent equations.
\begin{eqnarray}
\Sigma({\pmb p},\omega) & = & \frac{1}{N}\sum_{{\pmb q}}g^{2}({\pmb p},{\pmb q})
G({\pmb p}-{\pmb q},\omega-\Omega({\pmb q})) \nonumber \\
G({\pmb p},\omega) & = &
      \frac{1}{\omega-\varepsilon({\pmb p})-\Sigma({\pmb p},\omega)}
\end{eqnarray}
Here we use a magnon dispersion relation of the form 
$\Omega({\pmb q})=\sqrt{\Delta^{2}+v^{2}_{s}\left(1+\gamma({\pmb q})/4 \right)}$ with 
$\gamma({\pmb q})=2\big{(}\cos (q_{x})+\cos (q_{y})\big{)}$, which agrees in the 
limit ${\pmb q}\to{\pmb Q}=(\pi,\pi)$ with the form of Eq. (\ref{mdisp}).
Iterating the Green's function up to the 15th order to ensure convergence, 
we calculate the spectrum, 
$\rho({\pmb p},\omega)=\frac{1}{\pi}Im[G({\pmb p},\omega)]$ 
via the imaginary part of the Green's function 
and compute the
quasi-particle residue (QPR) at the first pole of the spectrum.
\begin{eqnarray}
\mathcal{Z}({\pmb p})=\Big{|}1-\frac{\partial}{\partial \omega}\Sigma'({\pmb
p},\omega)\Big{|}^{-1}_{\omega=\omega_{i}}
\end{eqnarray}
Figure \ref{scb1} shows the QPR for ${\pmb p}=(\pi,\pi)$ as a function of linear 
length $L$ of the square lattice for different values of the spin gap $\Delta$. 
The large-$L$ limit is indicated by a line.
Figure \ref{scb2} plots the quasiparticle weight as a function of the spin gap for
hole momenta ${\pmb p}=(\frac{\pi}{2},\frac{\pi}{2}),(0,\pi),(\pi,\pi)$. 

For hole  momenta satisfying $ \epsilon({\pmb p }) =  \epsilon({\pmb p } + {\pmb Q} ) $
( ${\pmb p}=(\frac{\pi}{2},\frac{\pi}{2})$ and 
${\pmb p}=(0,\pi)$ )   there is no energy denominator prohibiting the logarithmic 
divergence of the first order self-energy and  the QPR  shows an obvious decrease right  
up to a complete vanishing at the critical point.  Furthermore, the data is 
consistent with $\mathcal{Z} \propto \sqrt{\Delta}$. 
The case 
${\pmb p}=(\pi,\pi)$ 
is more complicated  since $ \epsilon({\pmb p }) \neq  \epsilon({\pmb p } + {\pmb Q} ) $. 
In first order, the self-energy remains bounded. The scattering of the hole of 
$ {\pmb Q}= (\pi,\pi) $  magnons leads to the progessive formation of 
shadow bands as  the critical point is approached  such that at  the critical 
point, the relation $ E^{(1)}({\pmb p})= E^{(1)} ({\pmb p}+{\pmb Q})$ holds.
This  back folding of the band 
can lead to the vanishing of the QPR  when higher order terms are included.   
Although the SCB results show a decrease of the QPR in the vicinity of the critical point, 
they are not accurate enough to answer the question of the vanishing of the QPR at this 
wave vector. 

\subsection{QMC approach}

\begin{figure}%[h]
%\centering
\includegraphics[height=4.5cm]{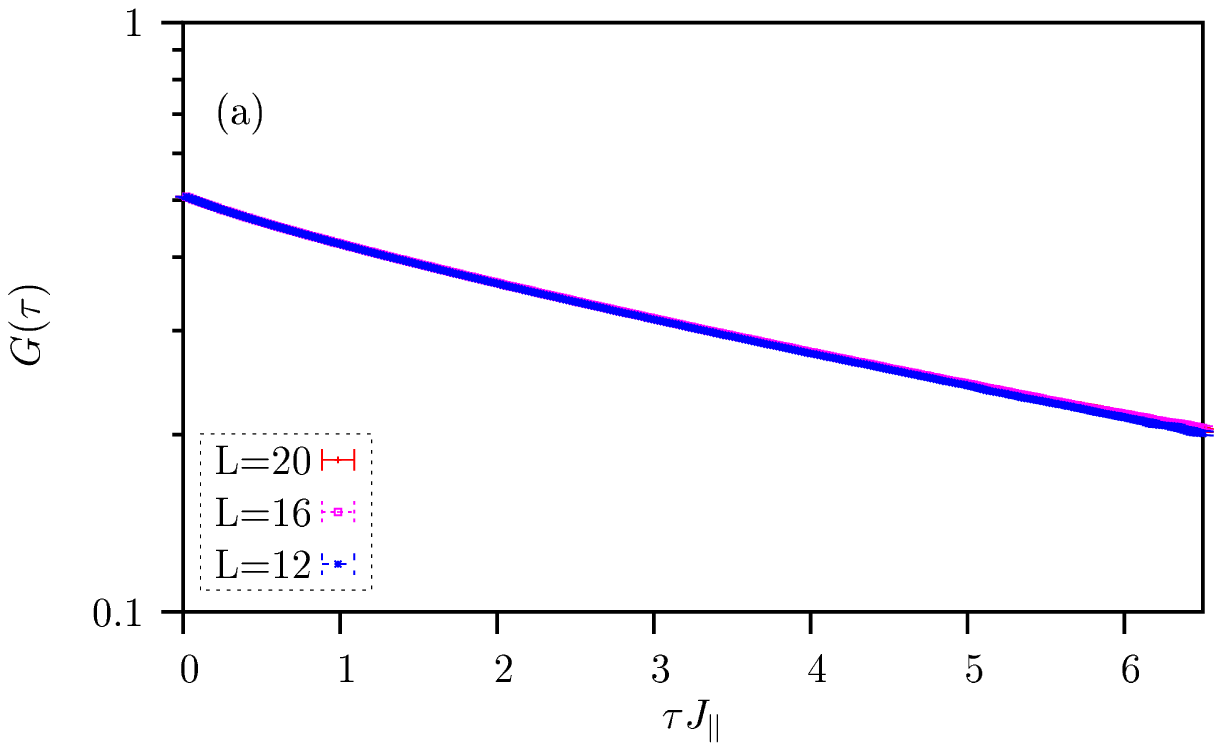}
\includegraphics[height=4.5cm]{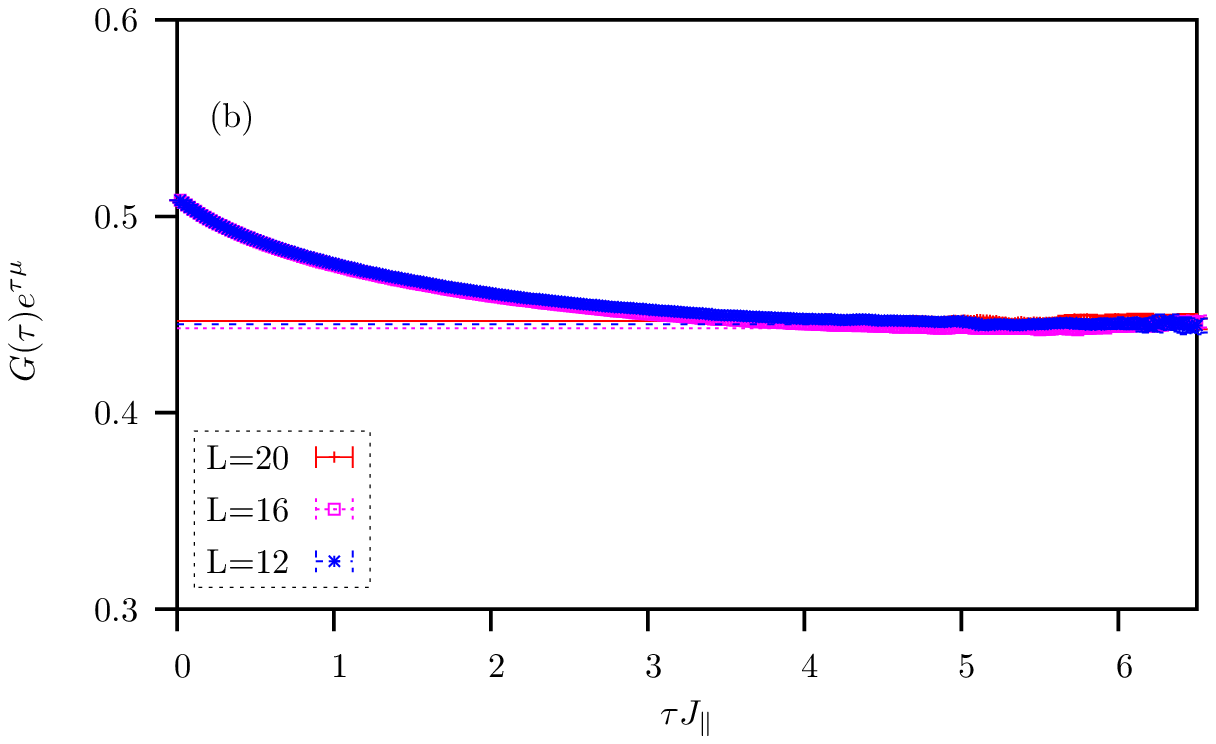}
\includegraphics[height=4.5cm]{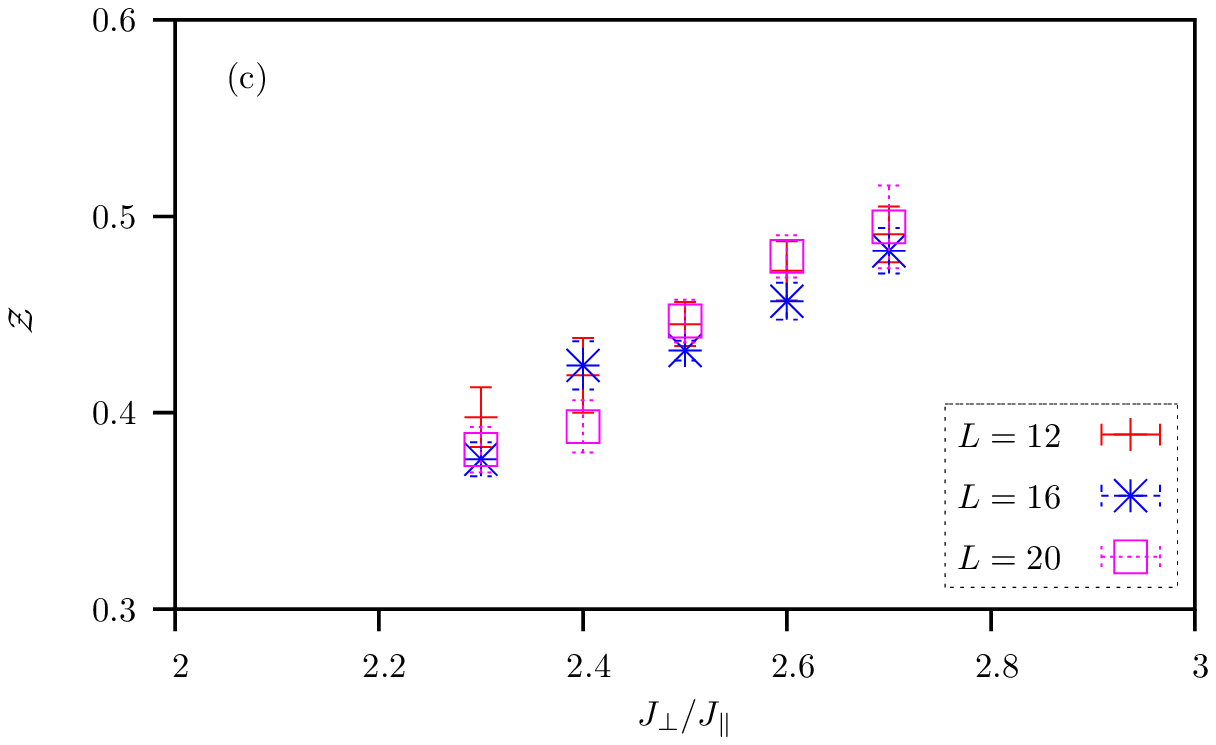}
\caption{\label{static_green}{\small Green's function in the vicinity of the phase transition 
($J_{\bot}/J_{\|}=2.5$) for a static hole and various  lattice sizes in the BHM 
(a) on a logarithmic plot and (b) on a plot 
where we adjusted the chemical potential in such a way that the Green's function converges 
to a constant value. Within the error bars and for lattice sizes greater then $12\times 12$ 
there is no size scaling recognisable. 
(Inverse temperatures: $\beta J_{\|}=30$ ($L=12$), $\beta J_{\|}=50$ ($L=16$)), 
$\beta J_{\|}=70$ ($L=20$); $\Delta\tau J_{\|}=0.02$)}
(c) QPR in the vicinity of 
the quantum critical point. }
\end{figure}

As shown in section \ref{methods}  we can extract the QPR 
from the asymptotic behavior of the Green's function. 
We first concentrate on the static hole in the BHM for which the QMC data is of higher 
quality that for the dynamic hole. Fig. \ref{static_green}a plots the Green's funtion as 
a function of lattice size at $ J_{ \bot } / J_{ \| } =2.5 $. As apparent within the 
considered range of imaginary times  no size and temperature effect is apparent.  We  fit the  
tail  ( $5 < \tau J_{\|} < 6$) of the Green's function to the form 
$\mathcal{Z} e^{-\tau \mu } $ and plot in Fig. \ref{static_green}b 
$G(\tau) e^{\tau \mu}$. In the large imgaginary time limit this quantitiy converges to the 
QPR $\mathcal{Z}$.  The so obtained value of $\mathcal{Z}$ is plotted for values of $J_{\bot} / J_{\|} $  
across the magnetic quantum phase transition. As apparent no sign  of the vanishing of 
the QPR is apparent as  we cross the quantum critcal point.

\begin{figure}%[h]
%\centering
\includegraphics[height=4.5cm]{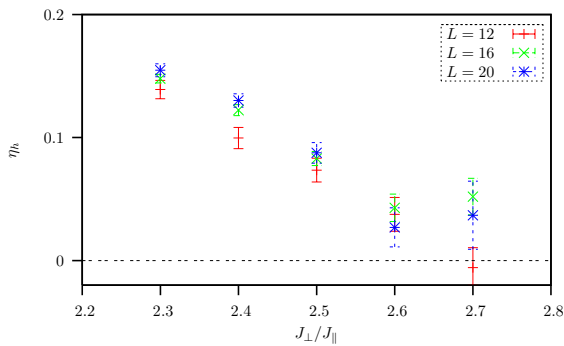}
\caption{ \label{eta_h_static.fig} { $\eta_h$ (see Eq. (\ref{greenfit}))  as a function of 
$J_{\bot}/J_{\|} $ for a static hole in the BHM. } }
\end{figure}

Our QMC data allows a different interpretation. 
Following the work of Sachdev et al. \cite{Sachdev01} we fit the imaginary time Green 
function to the form: 
\begin{eqnarray}
G(\tau) \propto \tau^{-\eta_{h}}\exp \left( -\tau \mu \right)
\label{greenfit}
\end{eqnarray}
in the the range $2.0< \tau J_{\|} <6.0$ as  done in Ref. \cite{Sachdev01}.  
Clearly, if $\eta_h > 0 $ then  the QPR vanishes. 
Our results for $\eta_h$ are plotted in Fig. \ref{eta_h_static.fig}.  
At $J_{\bot}/J_{\|}=2.5$  our result, $ \eta_h =0.0875\pm 0.0085 $ compares
very well to  that quoted in Ref. \cite{Sachdev01}, $\eta_{h}=0.087\pm 0.040$. 
The fact that the  result of Ref. \cite{Sachdev01} is obtained on a  $ 64 \times 64 $ lattice 
and ours on $20 \times 20$  confirms  that for the considered imaginary time 
range, size effects are absent.   Given the above interpretation of the data, Fig.
\ref{eta_h_static.fig} suggests that QPR of a static hole  vanishes 
for all $J_{\bot}/J_{\|} \leq  \left( J_{\bot}/J_{\|} \right)_c  \simeq 2.5 $. 

\begin{figure}[h]
%\centering
\includegraphics[height=4.5cm]{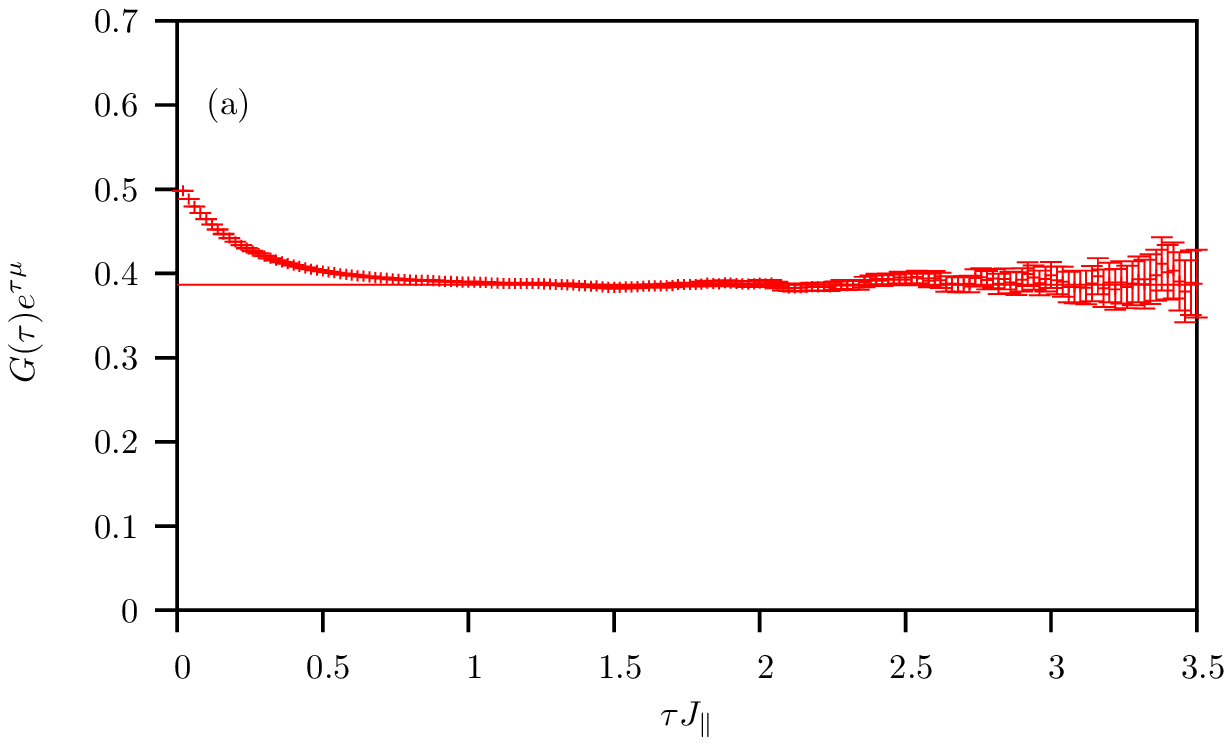}
\includegraphics[height=4.5cm]{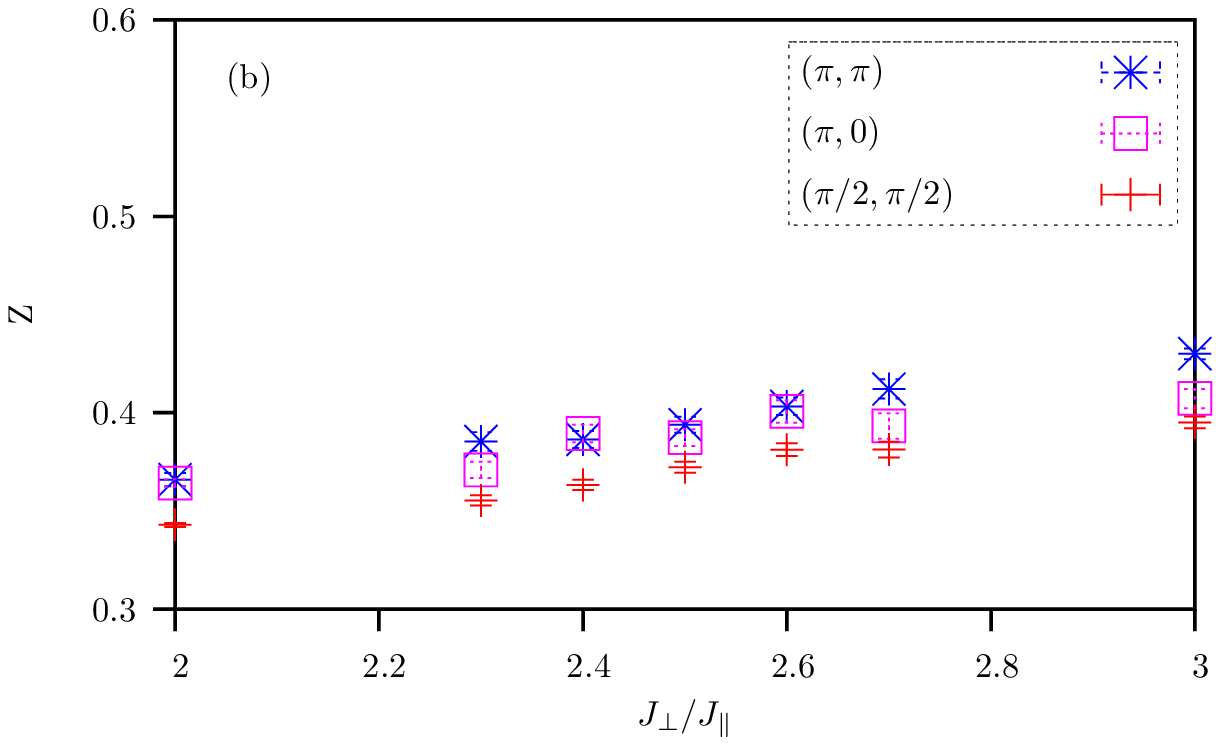}
\caption{\label{dynamic_green}{\small Green's function of a dynamic hole (${\pmb p}=(\pi,\pi)$) in 
the BHM for a $12\times 12$ lattice at  $J_{\bot}/J_{\|}=2.4$.} 
(b) QPR in the vicinity of 
the quantum critical point. }
\end{figure}

The choice of the fitting function reflects different ordering of the limits 
$\tau \rightarrow \infty $ and 
$ N \rightarrow \infty$.  On any finite size lattice the QPR is finite and 
hence it is appropriate to fit the tail of the Green's function to the form $\mathcal{Z}(N) e^{-\tau \mu}$
to obtain a size dependend QPR, and subsequently take the thermodynamic limit.  This strategy 
has been used successfully to show that the QPR of a doped  mobile hole in a one dimensional 
Heisenberg  chain vanishes \cite{Brunner00}.  On the other hand, the choice  of Eq. (\ref{greenfit})
for fitting the data implies that we first take the thermodynamic limit. Only in this limit, 
can the assymptotic form of the Green's function follow Eq. (\ref{greenfit}) with $ \eta_h \neq 0$. 
The fact that both procedures yield different results sheds doubt on the small imaginary 
time range used to extract the quasiparticle residue.  In particular, using data from 
$\tau J_{\|} = 2 $ onwards implies that we are  looking at a frequency window around the 
lowest excitation of the order $\omega / J \simeq 0.5$. Given this, it is hard to resolve the 
difference between a dense spectrum and  a well defined  low-lying quasiparticle  pole and 
a branch cut. 

\begin{figure}[h]
%\centering
\includegraphics[height=4.5cm]{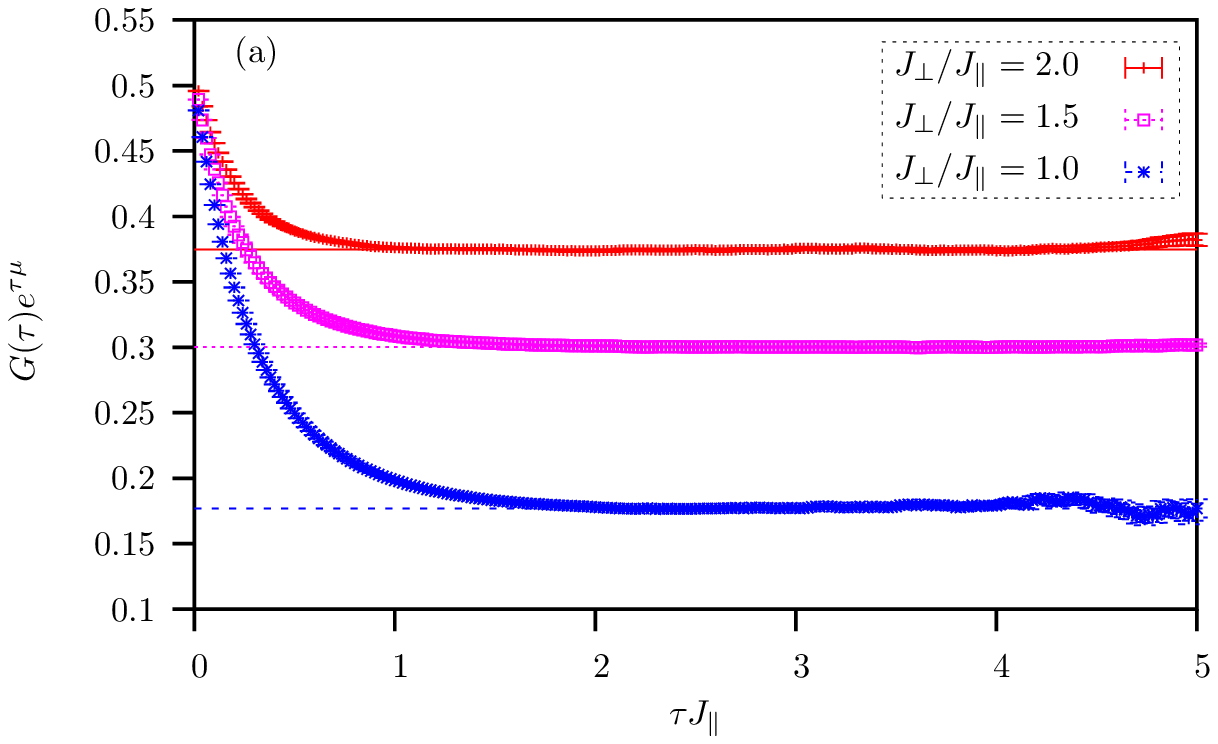}
\includegraphics[height=4.5cm]{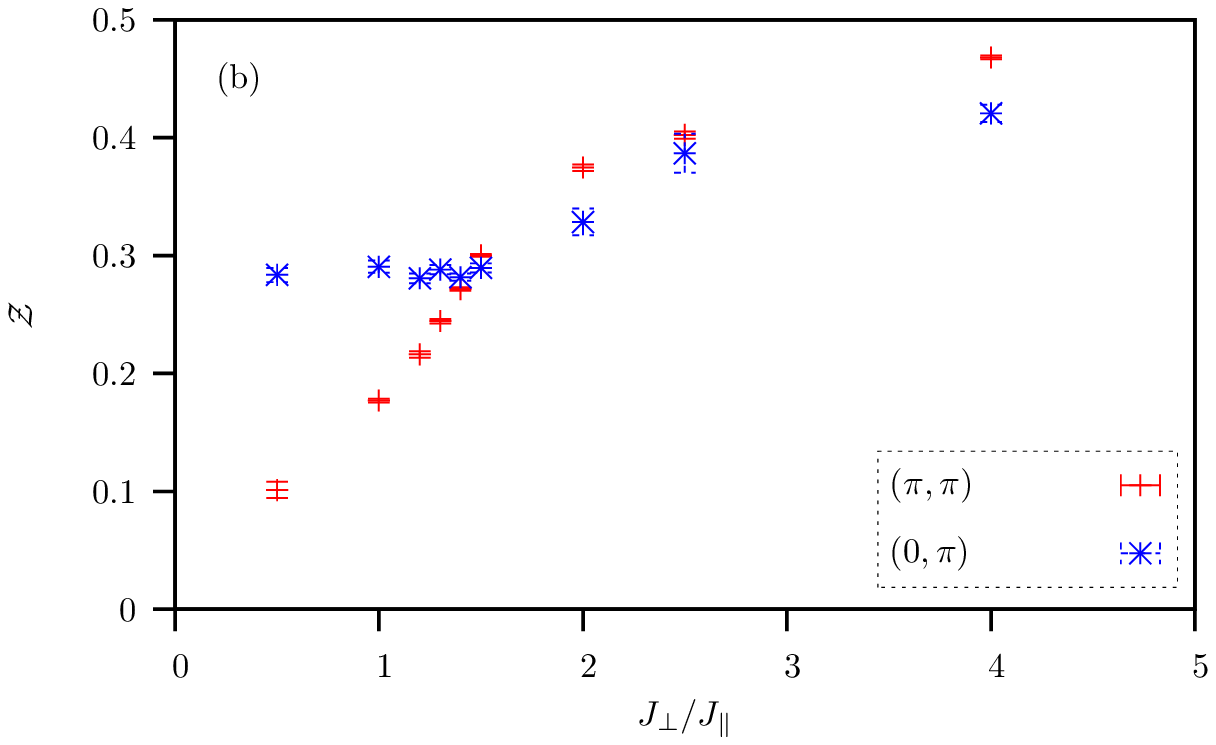}
\caption{\label{knm_qpr}{\small (a) Green's function $G_{\pmb p}(\tau)$ in the Kondo Necklace model 
at ${\pmb p}=(\pi,\pi)$ for $J_{\bot}/J_{\|}=1.0,1.5,2.0$. (b) Extracted values for the QPR. 
The critical point is localised at $(J_{\bot}/J_{\|})_{c}=1.360\pm 0.017$.}}
\end{figure}

We conclude this section by presenting data for a mobile hole in the BHM (see Fig. 
\ref{dynamic_green}) and  KNM (see Fig. \ref{knm_qpr}). Recall that in our simulations
we restrict the motion of the hole to a single plane. The data for the QPR in the above 
mentioned figures stem from fitting the tail of the Green's function to the 
form $\mathcal{Z}e^{-\tau \mu}$.  The fit to the form  of Eq. (\ref{greenfit}) yields values of $\eta_h$ 
which within the error bars are not distinguishable form zero. 

\section{Conclusion}
We have analyzed  single hole dynamics across magnetic order-disorder quantum phase 
transitions as realized in the Kondo Necklace and bilayer Heisenberg models.  The hole 
motion is restricted to the upper layer as appropriate  for interpretation of the data
in terms in Kondo physics.  Both models have identical spin dynamics since the
quantum phase transition is described by the $O(3)$ three-dimensional sigma  model 
\cite{Troyer97}.  On the other hand 
the single  hole dynamics shows marked differences.  In the strong coupling limit, deep in the 
disordered phase, the ground state of both models is well described by a direct product of 
singlets between the layers.  This Kondo screening  leads to a single hole dispersion relation
with maximum at ${\pmb p}=(\pi,\pi)$. In the Kondo Necklace model, where the spin degrees of
freedom on the lower layer interact indirectly through polarization of spin on the upper 
layer (RKKY type interaction),  the single hole dispersion  preserves it's maximum at 
$\pmb{p} =(\pi,\pi) $ down to arbitrarily low  interplanar couplings. This situation is 
very similar to the Kondo Lattice model of Eq. (\ref{KLM}). In this case, down to 
$\mathcal{J}/t = 0.2$,  substantially below the 
magnetic phase transition $\mathcal{J}_c/t = 1.45 \pm 0.05 $, the maximum of the the 
hole dispersion is 
pinned at $\pmb{p} = (\pi,\pi)$ and the effective mass at this ${\pmb p}$-point tracks the 
single ion  Kondo temperature \cite{Assaad04a}.  We note that this result is not 
supported by recent series expansions which show that there is a critical value of 
the coupling  where the effective mass diverges \cite{Trebst06}. Hence the 
interpretation that in both the Kondo 
necklace and Kondo lattice models, the localized spins remained partially screen down  to 
arbitrarily low values of the interlayer coupling. In other words, signatures of strong 
coupling physics in the single hole dispersion relation is present down to arbitrary low 
interplanar couplings.   

In the bilayer Heisenberg model where there is an independent energy scale coupling the 
spins on the lower layer, the situation differs.  At values of  $ J_{\bot} < J_{\bot,c}$  the 
maximum of the  single hole dispersion relation drifts towards $ {\pmb p} = (\pi/2,\pi/2) $ and 
the dispersion relation evolves continuously to that of a single hole doped in a planar 
antiferromagnetic  \cite{Brunner00b}.  Hence the interpretation that at weak couplings, 
Kondo screening in this model is completely suppressed.  In other words, the small but finite  
$J_{\bot}$ results can be  well understood starting form the $J_{\bot} =0 $ point.  

We have equally,  analyzed the quasiparticle residue across the magnetic order-disorder 
transition.  In the disordered phase using a bond mean-field approximation, there are two 
processes in which the hole couples to magnetic fluctuations (see Fig. \ref{coupling}): 
i) The hole propagates from one lattice site to another thereby rearranging the spin background.  
In the proximity of the 
critical point, and still within the bond-mean field approximation 
those processes do not couple to long range $\pmb{Q} = (\pi,\pi)$  magnetic fluctuations. 
A very similar result is obtained in the ordered phase \cite{Martinez91}.  ii)   In bilayer 
models  the hole can remain immobile and the spin in the lower layer can flip. Those processes
couple to critical magnetic fluctuations. Within a self-consistent Born approximation, this 
drives the quasiparticle residue to zero both for a static hole and mobile hole with momenta 
${\pmb p} $ satisfying $ \epsilon({\pmb p} + {\pmb Q}) = \epsilon({\pmb p})$. 
We have attempted  to confirm this point of view with Monte Carlo simulations. Within 
our quantum Monte Carlo approach,  where the accuracy of the single particle 
Green's function at large imaginary times is limited, we have found no convincing evidence
of the vanishing  of the quasiparticle  residue both for a static and a mobile hole. 
Furhter work and algortihmic developments are required to clarify this delicate issue. 

\vspace{0.5cm}
Acknowledgments.   The calculations were carried out on the Hitachi SR8000 of the LRZ M\"unchen. 
We thank  this institution for generous allocation of CPU time.  We have greatly profitted from
discussions with M. Vojta and L. Martin.  Financial support from the DFG under the grant number AS120/4-1 is 
acknowledged. 
%\bibliographystyle{./prsty}
%\bibliography{./bruenger,./fassaad}

\begin{thebibliography}{10}

\bibitem{Paschen04}
S. Paschen, T. L\"uhmann, S. Wirth, P. Gegenwart, O. Trovarelli, C. Geibel, F.
  Steglich, P. Coleman, and Q. Si, Nature {\bf 432},  881  (2004).

\bibitem{Schrieffer66}
J.~R. Schrieffer and P.~A. Wolff, Phys. Rev. {\bf 149},  491  (1966).

\bibitem{Tsunetsugu97}
H. Tsunetsugu, M. Sigrist, and K. Ueda, Rev. Mod. Phys. {\bf 69},  809  (1997).

\bibitem{Capponi00}
S. Capponi and F.~F. Assaad, Phs. Rev. B {\bf 63},  155114  (2001).

\bibitem{Feldbacher02}
M. Feldbacher, C. Jurecka, F.~F. Assaad, and W. Brenig, Phys. Rev. B {\bf 66},
  045103  (2002).

\bibitem{Brunner00}
M. Brunner, F.~F. Assaad, and A. Muramatsu, Eur. Phys. J. B {\bf 16},  209
  (2000).

\bibitem{Beach04}
K.~S.~D. Beach, cond-mat/0403055  (2004).

\bibitem{Sandvik98}
A. Sandvik, Phys. Rev. B {\bf 57},  10287  (1998).

\bibitem{Sushkov00}
O.~P. Sushkov, Phys. Rev. B {\bf 62},  12135  (2000).

\bibitem{Evertz97}
H.~G. Evertz, Adv. Phys. {\bf 52},  1  (1997).

\bibitem{Angelucci95}
A. Angelucci, Phys. Rev. B {\bf 51},  11580  (1995).

\bibitem{Troyer97}
M. Troyer, M. Imada, and K. Ueda, J. Phys. Soc. Jpn. {\bf 66},  2957  (1997).

\bibitem{Shevchenko00}
P.~V. Shevchenko, A.~W. Sandvik, and O.~P. Sushkov, Phys. Rev. B {\bf 61},
  3475  (2000).

\bibitem{Kotov98}
V.~N. Kotov, O. Sushkov, Z. Weihong, and J. Oitmaa, Phys. Rev. Lett. {\bf 80},
  5790  (1998).

\bibitem{Sachdev90}
S. Sachdev and R.~N. Bhatt, Phys. Rev. B {\bf 41},  9323  (1990).

\bibitem{Vojta99}
M. Vojta and K.~W. Becker, Phys. Rev. B {\bf 60},  15201  (1999).

\bibitem{Martinez91}
G. Mart\'{\i}nez and P. Horsch, Phys. Rev. B {\bf 44},  317  (1991).

\bibitem{Sachdev01}
S. Sachdev, M. Troyer, and M. Vojta, Phys. Rev. Lett. {\bf 86},  2617  (2001).

\bibitem{Assaad04a}
F.~F. Assaad, Phys. Rev. B {\bf 70},  020402  (2004).

\bibitem{Trebst06}
S. Trebst, H. Monien, A. Grzesik, and M. Sigrist, Phys. Rev. B {\bf 73},
  165101  (2006).

\bibitem{Brunner00b}
M. Brunner, F.~F. Assaad, and A. Muramatsu, Phys. Rev. B {\bf 62},  12395
  (2000).

\end{thebibliography}

\end{document}